\documentclass[a4paper,11pt]{article}
\usepackage{graphicx,amssymb,amstext,amsmath} 
\usepackage{color}

\definecolor{cbl}{rgb}{0,0,1}                

\newcommand{\bc}{\begin{center}}
\newcommand{\ec}{\end{center}}
\def\ba#1{\begin{array}{#1}\displaystyle}
\newcommand{\ea}{\end{array}}

\newcommand{\beq}{\begin{equation}}
\newcommand{\eeq}{\end{equation}}
\newcommand{\beqa}{\begin{eqnarray}}
\newcommand{\eeqa}{\end{eqnarray}}
\newcommand{\no}{\nonumber}
\newcommand{\n}{\nonumber\\}
\newcommand{\bi}{\begin{itemize}}
\newcommand{\ei}{\end{itemize}}
\def\mato#1{\left(\ba{#1}} 
\def\matf{\ea\right)}

\def\lt#1{\left#1}
\def\rt#1{\right#1}

\def\b#1{\bar{#1}}
\def\frc#1#2{\frac{#1}{#2}}

\newcommand{\p}{\partial}

\newcommand{\bra}{\langle}
\newcommand{\ket}{\rangle}
\newcommand{\Z}{{\mathbb{Z}}}
\newcommand{\N}{{\mathbb{N}}}
\newcommand{\R}{{\mathbb{R}}}
\newcommand{\C}{{\mathbb{C}}}

\newcommand{\uD}{{\mathbb{D}}}
\newcommand{\dd}{\mathrm{d}}
\newcommand{\ii}{{\mathrm{i}}}

\newcommand{\Or}{{\cal O}}

\newcommand{\ep}{\epsilon}
\newcommand{\varep}{\varepsilon}

\newcommand{\Tr}{{\rm Tr}}

\newcommand{\TT}{{\cal T}}

\newcommand{\rx}{{\rm x}}
\newcommand{\ry}{{\rm y}}

\newcommand{\cL}{{\cal L}}

\newcommand{\cV}{{\cal V}}
\newcommand{\ad}{{\rm ad}\,}

\newcommand{\PP}{{\mathbb P}}

\newcommand{\ri}{{\rm i}}

\begin{document}
\begin{titlepage}
\vspace{0.2cm}
\begin{center}

{\large{\bf{Conical Twist Fields and Null Polygonal Wilson Loops}}}

\vspace{0.8cm} 
{\large \text{Olalla A. Castro-Alvaredo${}^{\heartsuit}$, Benjamin Doyon{\LARGE${}^{\star}$}  and Davide Fioravanti{\LARGE{${}^{\bullet}$}}}}

\vspace{0.2cm}
{{\small ${}^{\heartsuit}$} Department of Mathematics, City, University of London, \\ 10 Northampton Square EC1V 0HB, UK }\\
\vspace{0.1cm}
{{\small {\LARGE{${}^{\star}$}}} Department of Mathematics, King's College London, Strand WC2R 2LS, UK}\\
\vspace{0.1cm}
{{\small {\LARGE{${}^{\bullet}$}}} Sezione INFN di Bologna, Dipartimento di Fisica e Astronomia, \\  Universit\`a di Bologna
Via Irnerio 46, Bologna, Italy}

\end{center}

\vspace{1cm}

Using an extension of the concept of twist field in QFT to space-time (external) symmetries, we study conical twist fields in two-dimensional integrable QFT. These create conical singularities of arbitrary excess angle. We show that, upon appropriate identification between the excess angle and the number of sheets, they have the same conformal dimension as branch-point twist fields commonly used to represent partition functions on Riemann surfaces, and that both fields have closely related form factors. However, we show that conical twist fields are truly different from branch-point twist fields. They generate different operator product expansions (short distance expansions) and form factor expansions (large distance expansions). In fact, we verify in free field theories, by re-summing form factors, that the conical twist fields operator product expansions are correctly reproduced. We propose that conical twist fields are the correct fields in order to understand null polygonal Wilson loops/gluon scattering amplitudes of planar maximally supersymmetric Yang-Mills theory.

\medskip

\vfill

\noindent 
${}^{\heartsuit}$ o.castro-alvaredo@city.ac.uk\\
{\LARGE{${}^{\star}$}} benjamin.doyon@kcl.ac.uk\\ 
{\LARGE{${}^{\bullet}$}} fioravanti@bo.infn.it
\hfill \today

\end{titlepage}
\section{Introduction}

\begin{figure}[h!]
\begin{center}
\includegraphics[scale=0.4]{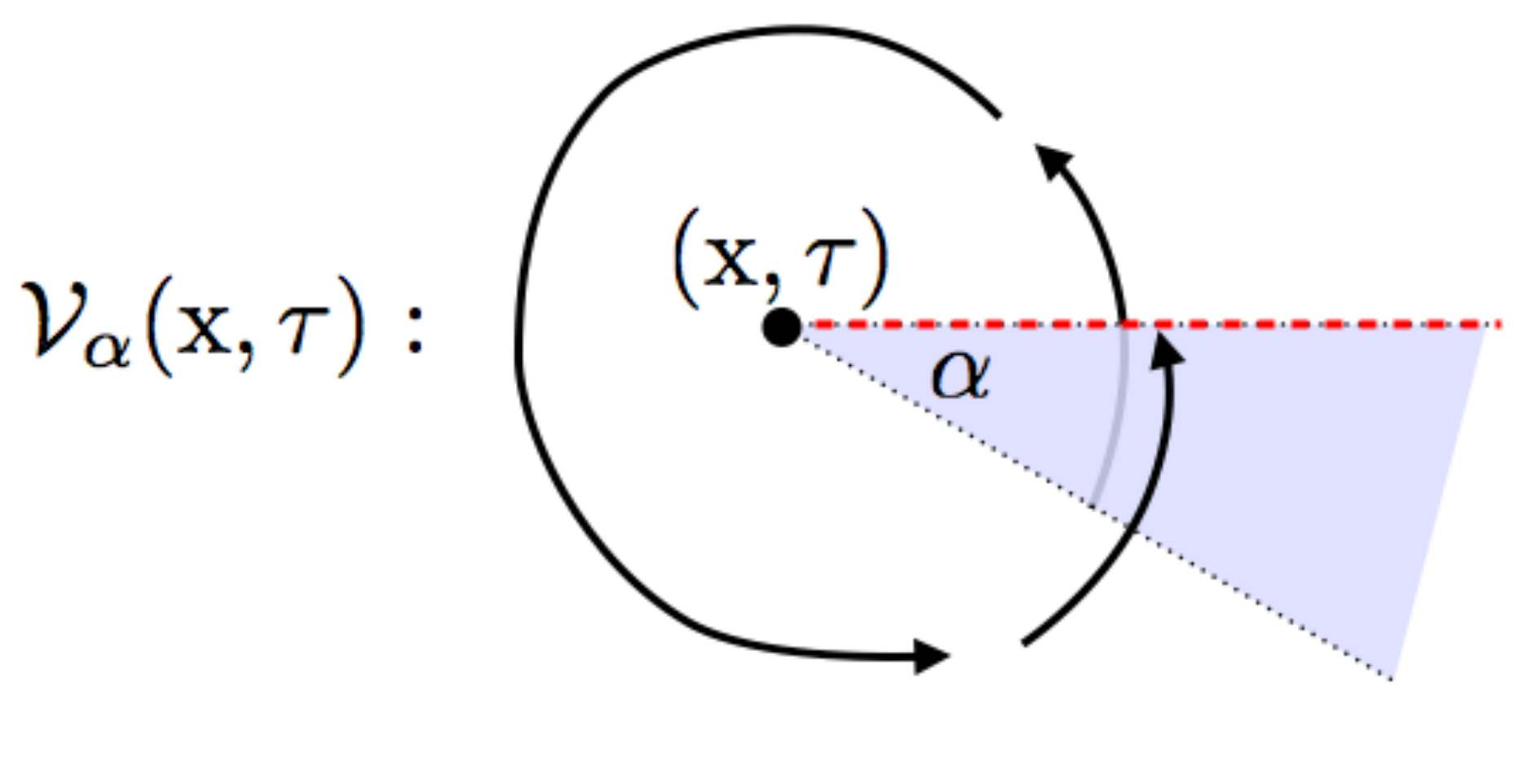} 
\caption{A pictorial representation of the conical twist field $\mathcal{V}_\alpha(\rx,\tau)$. The red horizontal line represents the branch cut induced by the conical singularity, and the arrows represent the continuous path a local field takes when continued around the position $(\rx,\tau)$. When reaching the cut, the local field is affected by a clockwise rotation of angle $\alpha$. Effectively, a wedge of angle $\alpha$ is added to the full cycle around the point $(x,\tau)$.}\label{figtwist}
\end{center}
\end{figure}

In quantum field theory (QFT) any singularity in an otherwise flat space-time is associated with a quantum field localized on the singularity. Since the stress-energy tensor $T^{\mu\nu}$ is the first order response to a small metric transformation, one may expect that metric singularities can be represented as exponentials of the stress-energy tensor. Natural examples of metric singularities are conical singularities in an otherwise (two-dimensional) Euclidean flat space: points of infinite curvature with excess angle $\alpha\neq 0$ (the curvature is positive (negative) for $\alpha<0$ ($\alpha>0$)). The goal of the present paper is to propose a quantum field, that we will call a {\it conical twist field}, expressed in terms of the stress-energy tensor, and that represents a generic conical singularity of excess angle $\alpha$. In the picture of Euclidean field theory, it is a field whose position $(\rx,\tau)$ is the end-point of a branch cut extending towards its right, through which other fields are affected by a rotation clockwise of angle $\alpha$ with respect to $(\rx,\tau)$. This is an extension of the well known concept of twist fields \cite{PhysRevB.13.316,Zuber:1976aa,1978NuPhB.144...80S,KW,sato1978holonomic,sato1978holonomicII,smirnovbook,Babelon:1992sn,Bernard:1992mu} to space-time symmetries, instead of internal symmetries. See Fig.~\ref{figtwist}.

Let $\frak R^\mu(y)$ (with $\mu=1,2$ representing space and imaginary time respectively)\footnote{Our convention is that if $\Or(x)$ is a spinless field, then $[\int_\R \dd \ry\,\frak R^2(\ry,\tau),\Or(\rx+\rx',\tau)] = \rx'\p_\tau \Or(\rx+\rx',\tau)$. In particular $\frak R^2(\ry,\tau)$ is hermitian.} be the conserved current associated to rotation symmetry with respect to the point $(\rx,\tau)$. The proposed conical twist field, positioned at $(\rx,\tau)$ and adding a wedge of angle $\alpha>0$ centered at $(\rx,\tau)$, is of the form
\beq\label{twr}
	\cV_\alpha(\rx,\tau)=\lt[e^{-\alpha\int_\rx^\infty \dd \ry\,\frak R^2(\ry,\tau )}\rt]
\eeq
where the square brackets indicate an appropriate renormalization of the field. We will define a renormalization procedure that makes the exponential finite and well defined, and show that the resulting field is a spinless scaling field with scaling dimension $2\Delta_\alpha$ given by
\beq
\Delta_\alpha=\frac{c}{24}\left(\frac{\alpha+2\pi}{2\pi}-\frac{2\pi}{\alpha+2\pi}\right).
\label{delta}
\eeq
This conical twist field has the unusual-looking hermiticity property
\beq\label{dag}
	\cV_\alpha^\dag = \cV_\alpha
\eeq
and generates the operator product expansion (OPE)
\beq
\mathcal{V}_\alpha (z)\mathcal{V}_{\alpha'}(0) =\mathcal{C}_{\alpha, \alpha'}^{\alpha+\alpha'} z^{-2\Delta_\alpha-2\Delta_{\alpha'}+2\Delta_{\alpha+\alpha'}} \mathcal{V}_{\alpha+\alpha'}(0)+ \cdots
\label{2}
\eeq
where $\mathcal{C}_{\alpha, \alpha'}^{\alpha+\alpha'}$ are the (yet unknown) three-point couplings.  In particular,
\beq
\mathcal{V}_{\alpha}(z)\mathcal{V}_{-\alpha}(0) = z^{-2\Delta_\alpha-2\Delta_{-\alpha}} \bf{1}+ \ldots
\label{4}
\eeq
where we take conformal field theory (CFT) normalization to fix the structure constant to $\mathcal{C}_{\alpha,-\alpha}^0=1$. Note that thanks to \eqref{twr}, conical twist fields can in principle be seen as exponentials of  Virasoro modes, or Virasoro vertex operators. Such exponentials implement conformal transformations and were studied in quite some generality in \cite{Bauer:2003kd}. In this sense, the conical twist field $\mathcal{V}_\alpha(0)$ implements the singular conformal transformation $z\mapsto z^{1/(1+\alpha/2\pi)}$.

In the study of QFT on branched Riemann surfaces, branch points of $n^{\rm th}$-root type, with integer $n$, are negative curvature conical singularities of excess angle $\alpha$ that are multiples of $2\pi$:
\beq\label{thetan}
	\alpha = 2\pi(n-1).
\eeq
Fields associated with conical singularities on Riemann surfaces have been studied a long time ago in the context of orbifold CFT \cite{kniz,orbifold}. More recently, this has been understood in general 1+1-dimensional QFT: such singularities can be studied using branch-point twist fields $\TT$ (and their conjugate $\tilde\TT$) \cite{entropy}. These are fields which exist in an $n$-copy replica model and which are associated to the (or any) generator of the $\Z_n$-symmetry of the replica model. The conical twist field is of a different (but related) nature, and does neither require $n$ to be an integer ($\alpha$ to be a multiple of $2\pi$) nor the model to consist of $n$ replicas (it exists in a single copy of the model). Its scaling dimension \eqref{delta} agrees with that of the branch-point twist field when $n$ is an integer and is related to $\alpha$ as in \eqref{thetan}. As we shall see in more detail later, its form factors agree with a particular subset of those of the branch-point twist field \cite{entropy} where all particles lie on the same copy. However, the hermiticity property \eqref{dag} and the OPE \eqref{2} differ from those of branch-point twist fields. Branch-point twist fields are generically not self-conjugate, and their OPEs do not involve branch-point twist fields at different values of $n$ (as $n$ is a property of the model, not of the field), but rather different elements of the $\Z_n$-symmetry. Further, the form factor series for the vacuum two-point function $\langle \TT(0) \tilde{\TT}(\ell)\rangle_n$ contains sums over replica indices, from 1 to $n$, as these parametrize the particles of the replica model. In contrast, that of conical twist fields does not involve such sums.
Conceptually, the difference between the conical twist field and the branch-point twist field is that the latter reproduces both a metric singularity and a particular branching (of $n^{\rm th}$-root type), while the former reproduces only the metric singularity, adding or deleting space with trivial branching. 

Partition functions on Riemann surfaces with branch points have received a lot of attention recently due to their use, within the ``replica trick'', for the evaluation of the von Neumann and R\'enyi entanglement entropies 
\cite{HolzheyLW94,Calabrese:2004eu,entropy} and the logarithmic negativity \cite{negativity1,negativity2,negativity3,ourneg}. In most of these applications, one establishes the result for integer $n$, and then performs an  analytic continuation to real values of $n$, which is in general hard to define uniquely and involves a certain amount of guesswork. However, in the case of the entanglement entropy for a single half-infinite interval in any QFT, one knows the exact reduced density matrix \cite{DM1,DM2,unruh}: it is the exponential of the integral of the rotation current, exactly as in \eqref{twr} (as a consequence of global conformal invariance, one then also knows it for a single interval of any length in CFT). In such cases the analytic continuation is simple, and this explains various aspects of the so-called entanglement spectrum \cite{ES}. Here, we study \eqref{twr} for the first time as a quantum field, discussing its OPEs and form factors.

Our motivation for the consideration of this field in part comes from recent works on gluon scattering amplitudes in planar ($N_c \rightarrow \infty$) $\mathcal{N} = 4$ Super Yang-Mills (SYM) theory, which are equal to expectation values of null polygonal Wilson loops (WLs) \cite{Alday-Malda}. In the simplest, non-trivial case, the six gluon amplitude, or equivalently the hexagonal WL, was tentatively identified by \cite{BSV}, in the strong coupling regime, with the form factor series for the two-point function $\langle \TT(0) \tilde{\TT}(\ell)\rangle_n$ of a branch point twist field $\TT$ and its hermitian conjugate $\tilde{\TT}$ in the massive $O(6)$ non-linear sigma model (NLSM). Strictly speaking, this applies to the contribution of the scalar excitations to the WL, where there are different expressions for all (particle) form factors\footnote{This applies, in most cases, to the sum on $O(6)$ internal indices of the form factor modulus squares.} \cite{BSV, DF1, DF4, Bel}, although their detailed derivation is still missing. {\it A posteriori}, this identification was supported by the fact that a short-distance re-summation of the series reproduces the right conformal limit and scaling dimension for the twist field \cite{BSV, DF1, DF4, Bel} (with different computation strategies, as mentioned below).

However, the form factor series of $\langle \TT(0) \tilde{\TT}(\ell)\rangle_n$, as first obtained in \cite{entropy}, is in fact, strictly speaking, {\em different} from that found for six gluon amplitude or hexagonal WL. Indeed, as mentioned, the form factor expansion of $\langle \TT(0) \tilde{\TT}(\ell)\rangle_n$ contains sums over replica indices. In contrast, six gluon amplitudes or hexagonal WL, although formed out of the same building blocks (modulus square of twist field form factors) \cite{BSV}, do not contain replica sums. Furthermore, in the amplitude/WL case the value of $n$ is set to $n=\frac{p}{4}=\frac{5}{4}$ -- through a pictorial analogy with a pentagon, where the particle jump $p=5$ times on the edges to go back to its starting point, into which the hexagon is decomposed. Yet $n$ must be an integer in order for branch-point twist fields to be defined, as it counts the number of replicas; non-integer values of $n$ are obtained by analytic continuations and don't strictly correspond to two-point functions.

Instead, we will argue that six gluon amplitudes/hexagonal WLs may be expressed as two-point functions $\langle \mathcal{V}_\alpha (x)\mathcal{V}_{\alpha}(0)\rangle$ of the {\it conical} twist fields introduced above, with $\alpha=\pi/2$. Indeed, the form factor series for $\langle \mathcal{V}_\alpha (x)\mathcal{V}_{\alpha}(0)\rangle$ is in general of the same form as the series describing the six gluon amplitude or hexagonal WL, in particular without sum over replica index. In addition, as per the prescription for the form factors of $\mathcal{V}_{\alpha}$, the form factors in the series for $\langle \mathcal{V}_{\pi/2} (x)\mathcal{V}_{\pi/2}(0)\rangle$ coincide with those of the twist field $\TT$ where all particles are on the same copy for the choice $n=\frac{5}{4}$, in agreement with the amplitude/WL result. In \cite{BSV} an OPE was identified, which can be written as
\beq
\mathcal{V}_{\frac{\pi}{2}}(z)\mathcal{V}_{\frac{\pi}{2}}(0) \sim z^{-4\Delta_{\frac{\pi}{2}}+2\Delta_{\pi} }\Or_{\rm hexa}+\cdots.
\label{ori}
\eeq
In this OPE, $\Or_{\rm hexa}$ was interpreted as associated to an ``hexagon" amplitude with form factors equal to those of the branch point twist field $\TT$ for $n=\frac{3}{2}$ (again with all particles on the same copy). This leads to the identification $\Or_{\rm hexa}=\mathcal{V}_\pi$, which again agrees with the general conical field OPE \eqref{2}, and further supports our conjecture. The power in (\ref{ori}) is
$
4\Delta_{\frac{\pi}{2}}-2\Delta_{\pi}=\frac{c}{180}.
$
In particular, for $c=5$ as in the $O(6)$ NLSM, this gives exponent $\frac{1}{36}$ which has been recovered very precisely by re-summing the form factor expansion in \cite{BSV, DF1, DF4, Bel}. In \cite{DF1,DF4} this was done by applying to the asymptotically free $O(6)$ NLSM the well-known cumulant expansion \cite{Smir} reviewed and employed below.

In other words, with the definition given here of the conical twist field we have set on solid ground the interpretation of the form factor series obtained for gluon scattering amplitudes/null polygonal WLs. The exact interpretation was made in the massless limit of the $O(6)$ NLSM, realised by the infinite 't Hooft coupling limit. But, as supported in the following by the form factor properties we derive, we would expect this identification to be true for the massive $O(6)$ NLSM realised by the strong coupling regime, and similar conical fields to be involved in the full string NLSM replacing the $O(6)$ NLSM (\cite{AM} and references therein) at any coupling. Further, one can generalise to other polygons: in the WL picture, $k$-sided polygonal Wilson loops correspond to $(k-4)$-point correlation functions (the heptagon corresponds to the three point function, the octagon to the four point function, and so on). Finally, for the full theory we expect that the conical twist field form factors are associated to an integrable theory with scattering matrix defined on the GKP vacuum \cite{FPR1,BR,DF2}.  

The paper is organised as follows: in section 2 we provide a rigorous QFT definition of conical twist fields and discuss their properties at and near conformal critical points. In particular we argue that, unlike the more standard definition of a twist field in QFT as a field associated with an internal symmetry of the theory, conical twist fields are associated with rotational space-time symmetry (hence an external symmetry) and insert conical singularities corresponding to excess rotation angle $\alpha$. They may therefore be formally expressed as (appropriately regularised) exponentials of integrals of the rotation Noether current over the line that starts at the twist field insertion point and extends up to infinity. In section 3 we characterise the same fields through their form factors and write the corresponding form factor equations for integrable 1+1 dimensional QFTs with  diagonal scattering. We give the general form of the two-particle form factor and, for free theories, give also closed formulae for all higher particle form factors. In section 4 we review the standard cumulant expansion of two-point functions of local fields in terms of their form factors and specialize it to conical twist field two-point functions in free theories. For free theories we obtain an exact resummation of the leading short-distance contribution to the correlators. In section 5 we test our form factor formulae by numerically evaluating our analytic expressions for the leading short-distance behaviour of the correlator $\bra \mathcal{V}_\alpha(0)\mathcal{V}_{\alpha'}(\ell)\ket$ and finding that they exactly match the expected decay from the conformal OPEs. In particular, we observe that the form factor resummation becomes subtle when any of the excess angles is negative in which cases a detailed analysis of the integrands' pole structure is required. This is reminiscent of the kind of issues that arise when analytically continuing the correlators of branch point twist fields from $n$ integer to $n\geq 1$ and real. We present our conclusions in section 6. A  derivation of four defining properties of the conical twist field is presented in appendix A. 

\section{Conical twist fields in quantum field theory}

The construction of the conical twist field is based on the standard theory of twist fields associated with internal QFT symmetries. Recall that in the quantization on the line, a twist field $\cV_\sigma$, associated with an internal symmetry $\sigma$ of the QFT, satisfies equal-time exchange relations
\beq\label{exch}
	\cV_\sigma(\rx,\tau)\, \Or(\ry,\tau) = \lt\{\ba{ll}
	\sigma\cdot \Or(\ry,\tau) \,\cV_\sigma(\rx,\tau) & (\ry>\rx) \\
	\Or(\ry,\tau) \,\cV_\sigma(\rx,\tau) & (\ry<\rx)
	\ea\rt.
\eeq
where $\sigma\cdot\Or$ is the transformation of $\Or$ under $\sigma$.
Through the usual time-ordering prescription, a correlation function with a twist field insertion may be evaluated by a path integral over field configurations with a jump condition across the cut $\{(\ry,\tau):\ry>\rx\}$. The jump condition imposes continuity between any local field $\Or$ just below the cut, and its transform $\sigma\cdot\Or$ just above the cut. Since $\sigma$ is a symmetry, the result only depends on the homotopy class of the cut on space-time with punctures at the positions of other local fields in the correlation function. A twist field is local in the sense that it commutes at space-like distances with the stress-energy tensor, as the latter is invariant under $\sigma$. 

Let us extend this concept to actions of $\sigma$ on space-time (external symmetries). Specifically, let $\sigma$ be a {\em rotation clockwise} by an angle $\alpha>0$ with respect to the point $z=\rx+\ii \tau$. For instance, if $\Or(\ry,\tau)$ has spin $s$, and denoting the position by a complex coordinate $w = \ry+\ii\tau$, we have $\sigma\cdot\Or(w) = e^{-\ii s\alpha} \Or(e^{-\ii\alpha}w)$\footnote{This operator is not necessarily holomorphic but may depend also on $\bar w$ which transforms accordingly.}. Let us denote by $\cV_\alpha(\rx,\tau)$ the corresponding conical twist field satisfying the exchange relations \eqref{exch}\,\footnote{Relation \eqref{exch} defines a family of twist fields.  The field $\cV_\alpha$ is further defined by imposing appropriate minimality conditions, such as minimality of its scaling dimension, and is here assumed to be unique. Recall that such conditions also play a role in the identification of $\TT$ and $\tilde{\TT}$ as the correct branch point twist fields.}. The associated jump condition in the path-integral representation imposes continuity between fields just below the cut $\{(\ry,\tau):\ry>\rx\}$ and clockwise $\alpha$-rotated fields just above the cut. Intuitively, a clockwise rotation  brings the fields from just above the half-line to a rotated position in an ``extra'' space-time wedge below the half line, thus adding a wedge centered at $(\rx,\tau)$ of angle $\alpha$ and creating a negative-curvature conical singularity. See Fig.~\ref{figtwist}.

Since rotations form a Lie group, they have an associated Noether current $\frak R^\mu(y)$. This immediately leads to the exponential representation of $\cV_\alpha(\rx,\tau)$ in terms of the stress-energy tensor, as in \eqref{twr}. Indeed, the standard commutation relations between the rotation Noether current and local fields then guarantees that this exponential generates the correct exchange relations \eqref{exch} with $\sigma$ the clockwise $\alpha$-rotation. Clearly, conical twist fields are {\em not} local, even in the large sense of commuting with the stress-tensor at space-like distances. However, since ${\frak R}^\mu$ is a conserved current, the field $\cV_\alpha(\rx,\tau)$ does not depend on the shape of the cut emanating from its position, but, again, only on its homotopy class on space-time with punctures at the positions of inserted local fields. This is the sense in which $\cV_\alpha(\rx,\tau)$ may be understood as enjoying a certain locality property. Hence we have more generally
\beq\label{twre}
	\cV_\alpha(\rx,\tau)=\lt[e^{-\alpha\int \dd s\, \frac{\dd y^\mu(s)}{\dd s}\, \ep_{\mu\nu}\,\frak R^\nu(y(s))}\rt]
\eeq
for any curve $s\mapsto y^\mu(s)$ connecting $(\rx,\tau)$ to $\infty$.

In the rest of this section, we make these ideas more precise and state the main properties of the resulting field.

%
%

\subsection{Rotation current and singular curvature}

In the following, it will be convenient to use complex coordinates $z=\rx + \ii\tau$, $\b z = \rx-\ii\tau$ and the  stress-energy tensor components $T=-(\pi/2)\, T^{\b z\b z}$ and $\b T =-(\pi/2) \,T^{zz}$ with the usual CFT normalization, $T(z)T(z')\simeq (c/2)\,(z-z')^{-4}$, $\b T(\b z)\b T(\b z')\simeq (c/2)\,(\b z-\b z')^{-4}$ where $c$ is the central charge of the ultraviolet CFT. We will also denote by $\Theta = -(\pi/2) \,T^\mu_\mu$ the ``CFT-normalized'' trace of the stress-energy tensor. In particular, we have $\b\p T + \p \Theta=0$ and $\p \b T + \b\p\Theta =0$.

For simplicity we concentrate on the conical twist field $\cV_\alpha:=\cV_\alpha(0,0)$ positioned at the origin. Note that in order for the expression \eqref{twr} to generate the correct exchange relations, only the commutators of the rotation current $\frak{R}^\mu$ with other local fields are required. Therefore, the choice of the current $\frak{R}^\mu$ defining the conical field is ambiguous with respect to addition of terms proportional to the identity operator ${\bf 1}$. This affects the normalization of the field $\cV_\alpha$. We propose to lift the ambiguity by adopting the following rotation current:
\beqa
	\ii\pi \frak R^z &=&
	\b z \b T - z \Theta - \frc{1}{\b z} {\cal G}(m |z|){\bf 1} \n
	-\ii\pi \frak R^{\b z} &=&
	z T - \b z \Theta - \frc{1}{z}{\cal G}(m |z|){\bf 1}
	\label{r0}
\eeqa
where $m$ is a mass scale of the QFT model and ${\cal G}(r)$ with $r:=m |z|$ is a universal scaling function with
\beq\label{G0}
	{\cal G}(0)=\frc{c}{12}.
\eeq
At the conformal point, where there is no mass scale, the factor ${\cal G}(r)$ is the constant $c/12$ in \eqref{r0}. Except for the singularity at the origin in \eqref{r0}, given by \eqref{G0}, the function ${\cal G}(r)/r$ is required to be integrable on the line. Its exact form does not affect the evaluation of correlation functions, but might be involved in the evaluation of vacuum expectation values of conical fields in massive QFT. 

Although we will not fully address the problem of the exact form of ${\cal G}(r)$ here, we nevertheless propose that the function ${\cal G}(r)$ should take the form
\beq\label{sca}
	{\cal G}(r) = \frc{c}{12} - 2r^2 g(r) + 2\int_0^r \dd r'\,r'g(r'),\quad
	g(r) = m^{-2} \bra\Theta(x)\ket_{\C,K}
\eeq
where the expectation value $\bra\Theta(x)\ket_{\C,K}$ is evaluated on the plane $\C$ with a Gaussian curvature $K$ that is {\em singular at the position of the conical field},
\beq
	K(x) = -2\pi \delta^{(2)}(x).
\eeq

The singularity in \eqref{r0}, with residue specified by \eqref{G0}, is essential in order to guarantee the correct scale transformation properties of the conical field. We provide in Appendix \ref{appcon} a CFT calculation that illustrates this.  A further justification for this singularity, as well as for the choice \eqref{sca} beyond CFT, is as follows. At the insertion of the conical field, a conical singularity emerges. Singularities in QFT give rise to additional renormalizations, and at the singular point microscopic effects are important. In order to account for these, a standard regularization is performed by making a hole in space-time around the singularity (see below). The limit is then taken in which the hole is made infinitesimally small. At the boundary of the hole, the trace of the stress-energy tensor acquires a nonzero expectation value proportional to the (negative) linear curvature. The boundary integral of this linear curvature is $-2\pi$. In the limit where the hole becomes a point, this may be replaced (in accordance with the Gauss-Bonnet theorem) by a punctual singular Gaussian curvature $K(x)=-2\pi \delta^{(2)}(x)$. We account for this by shifting the stress-tensor trace by its expectation value on the space $\C$ with the singular curvature $K$,
\beq
	{\tt\Theta} := \Theta - \bra\Theta(x)\ket_{\C,K} {\bf 1}.
\eeq
The shift has a delta-function contribution at the origin, which is inferred from the CFT trace anomaly formula
\beq
	\bra\Theta(x)\ket_{\C,K}\stackrel{CFT}= -\frc{cK(x)}{24}.
\eeq
The other elements of the stress-energy tensor, $\frak T$ and $\b{\frak T}$, are shifts of $T$ and $\b T$ respectively, determined by the conservation equations $\b\p \frak T + \p \tt \Theta=0$ and $\p {\b {\frak T}} + \b\p \tt\Theta=0$. Accounting for the delta-function singularity using $\b\p (1/z)= \p(1/\b z) = \pi\delta^{(2)}(x)$, we then obtain 
\[
	{\frak T} = T-\frc{c}{12 z^2} f(r){\bf 1},\quad
	\b{\frak{T}} = \b T-\frc{c}{12 \b z^2} f(r){\bf 1},\quad
	{\tt \Theta} = \Theta+\frc{cm^2 }{12} h(r){\bf 1}
\]
where $(cm^2/12)h(r) = -\bra\Theta(x)\ket_{\C,K}$ and $f(0)=1$ (and $\lim_{r\to0} h(r)<\infty$). The conservation equations imply that, away from the origin, $f$ and $h$ satisfy $r^2h'(r) = f'(r)$. The rotation current takes the standard form using these shifted stress-energy tensor components, $\ii\pi \frak R^z =\b z \b {\frak T} - z \tt\Theta$ and $-\ii\pi \frak R^{\b z} =  z \frak T - \b z \tt\Theta$, which gives the universal scaling function ${\cal G}(r) = (c/12)(f(r)+r^2h(r))$ in agreement with \eqref{sca}.

Note that the integration measure involved in \eqref{twre} is $\dd x^\mu\ep_{\mu\nu}\frak R^\nu(x) = (\dd z\,\frak R^z - \dd \b z\,\frak R^{\b z})/(2\ii)$, where $\ep_{\mu\nu}$ is the anti-symmetric symbol. In CFT, this specializes to
\beq\label{r0cft}
	\dd x^\mu \ep_{\mu\nu} \frak R^\nu(x) = -\frc{1}{2\pi }\lt(z T\, \dd z + \b z \b T\, \dd\b z - 
	\frc{c}{12} \lt(\frc{\dd z}z + \frc{\dd \b z}{\b z}\rt){\bf 1}\rt) \qquad\mbox{(CFT).}
\eeq

In the next subsection we provide further calculations showing that the choice of rotation current  \eqref{r0cft} is correct in CFT. Away form the conformal point, the expression \eqref{r0} with \eqref{sca} constitutes a conjecture.

\subsection{The conical field}\label{ssectsingle}

Besides the choice of the current, the right-hand side of \eqref{twr} requires a regularization / renormalization procedure. The theory developed in \cite{Bauer:2003kd} allows in principle to regularize such Virasoro vertex operators by separating the exponential into a product of exponentials with positive and negative Virasoro generators. However, making the full connection with \cite{Bauer:2003kd} is beyond the scope of this paper, and instead we concentrate on a more direct renormalization, if more difficult to control. We use an ``angular renormalization'' procedure based on an angle-splitting prescription. This takes inspiration from \cite{freefield1,freefield2} where angular quantization was used in order to study $U(1)$ twist fields in the Dirac theory. Here no explicit quantization scheme is used and we concentrate on correlation functions.

We first define a regularized exponential of a line integral starting at 0 in the complex plane. We regularize the Taylor expansion as follows: (a) a circular hole of radius $\varep>0$ is made around the origin, on the boundary of which we impose conformal boundary conditions, modifying the lower limit of the integral to a position on this boundary; (b) powers of the integral are split into products of integrals along rays at different angles. We then take the limit where all angles are equal to each other. That is, an $\varep$-regularized exponential is defined, inside any correlation functions on an open set $D\ni\{0\}$, as
\beq\label{regang}
	\lt\bra\lt[e^{\int_{0}^{L} \dd  x^\mu f_\mu(x)}\rt]_\varep\cdots\rt\ket_D =
	\sum_{j=0}^\infty \frc{1}{j!}
	\lim_{\{\phi_k\to0\}}\;
	\lt\bra\prod_{k=1}^j \;
	\int_{\varep e^{\ii\phi_k}}^{Le^{\ii\phi_k}} \dd x^\mu f_\mu(x)
	\;\cdots\rt\ket_{D\setminus \varep \uD}
\eeq
where $L>\varep$ and we use complex numbers in order to represent coordinates in the integration limits. Here the ellipsis $\cdots$ represent potential insertions of local fields at fixed positions.

In order to define the conical twist fields, we then take a renormalized limit where $\varep\to0$ of the regularized exponential. On the plane, the field corresponding to the insertion of a single conical singularity of excess angle $\alpha$ at the origin is:
\beq\label{renang}
	\cV_\alpha:= \lim_{\varep\to0} \varep^{-2\Delta_\alpha}
	\lt[e^{- \alpha\int_0^\infty \dd x^\mu \ep_{\mu\nu}\frak R^\nu(x)}\rt]_\varep
\eeq
where $\Delta_\alpha$ is given by \eqref{delta}. In CFT this suffers from infrared divergences. One may regulate these infrared divergencies by a finite-volume cutoff. For instance, on the disk $\ell\uD$ of radius $\ell$, we define the conical field at position 0 as
\beq\label{renang2}
	\cV_\alpha\big|_{\ell\uD}:= \lim_{\varep\to0} \varep^{-2\Delta_\alpha}
	\lt[e^{- \alpha\int_0^\ell \dd x^\mu \ep_{\mu\nu}\frak R^\nu(x)}\rt]_\varep.
\eeq
The latter is in fact valid both in massive QFT and in CFT, and the limit $\ell\to\infty$ can be taken in massive models and reproduces \eqref{renang}.

Recall that ${\frak R}^\mu$ is a conserved Noether current associated to the rotation symmetry. Therefore, the integral $\int_{\varep e^{\ii\phi}}^{\ell e^{\ii\phi}} \dd x^\mu \ep_{\mu\nu} {\frak R}^\nu(x)$, inside correlation functions on the annulus of inner radius $\varep$ and outer radius $\ell$ with conformal boundary conditions, is invariant under a change of the angle of the path $\phi$, as long as the path does not cross the positions of other local fields inserted. As a consequence, the limits on the angles $\phi_k$ in \eqref{regang}, on the right-hand sides of \eqref{renang} and \eqref{renang2}, exist: the correlation functions are independent of the angles $\phi_k$ if they are near enough to each other.

Note that, again using current conservation, it is not necessary to take paths that are straight rays: for the purpose of the conical twist field, the regularized exponential can be defined by taking, for the $k^{\rm th}$ term of the Taylor expansion, any collection of $k$ non-crossing paths from one boundary to the other of the annulus, and by taking the limit, order by order, where they accumulate to a single ray from $\varep$ to $\ell$.

In Appendix \ref{appcon}, we provide field theory arguments for the following statements. Point II is expressed here, and shown in Appendix \ref{appcon}, in the case $\alpha\in[0,2\pi)$ for simplicity.
\bi
\item[I.] The limit $\varep\to0$ in \eqref{renang} and \eqref{renang2} exits and is finite inside correlation functions.
\item[II.] The field $\cV_\alpha$ (on the disk or the plane) implements a conical singularity at the origin of angle $2\pi+\alpha$: it makes a cut on ${\cal L}=(0,\ell)$ and inserts there a wedge in a disjoint copy of $\ell\uD$ bounded by the segments ${\cal L}_+=(0,\ell)$ and ${\cal L}_-=e^{-i\alpha}(0,\ell)$, by establishing at $\cal L$ continuity from below to ${\cal L}_-$, and from above to ${\cal L}_+$ (take $\ell\to\infty$ for the plane).
\item[III.] The field $\cV_\alpha$ (on the disk or the plane) behaves as a spinless scaling field at position 0 with scaling dimension $2\Delta_\alpha$ given by (\ref{delta}).
\item[IV.] The field is conformally normalized, that is
\beq\label{norm}
	\bra \cV_\alpha\ket_{\ell\uD} = \ell^{-2\Delta_\alpha}\qquad
	\mbox{(CFT)}.
\eeq
\ei
Note that Properties II and III imply the OPEs \eqref{2} (and \eqref{4}): as two conical fields approach each other, their associated excess angles must add up, and the ensuing singularity 
is determined by the scaling dimensions of the fields involved.

Let us be more precise concerning Property II. Recall that we may construct a manifold ${\cal M}_{\alpha,\ell}$ with a conical singularity as follows; for simplicity we restrict our attention to $\alpha\in(0,\pi)$. We consider the generic case of a disk $\ell\uD$, with rotation-invariant boundary conditions on the boundary of the disk. The result holds as well on the plane by taking $\ell\to\infty$, with $(\ell\uD)_{\ell\to\infty}$ identified with the plane $\C$. Let $\ell\uD^{(j)}$, $j=0,1,2$ be three copies of the disk. Consider two copies of the disk minus segments, $\check\uD^{(1)} = \ell\uD^{(1)}\setminus[0,\ell)$ and $\check\uD^{(2)} = \ell\uD^{(2)}\setminus e^{-i\alpha}[0,\ell)$, and a copy of the punctured disk $\check\uD^{(0)} = \ell\uD^{(0)}\setminus\{0\}$. For any $0\leq a<b\leq 2\pi$, denote the subsets $\{z\in \check\uD^{(j)}:{\rm arg}(z)\in(a,b)\}$ by $\check\uD^{(j)}_{(a,b)}$. The manifold can be covered by three patches: the two cut disks $\check\uD^{(1)}$ and $\check\uD^{(2)}$, and a wedge $\check\uD^{(0)}_{(-\alpha,\alpha)}$. The transition functions between these patches are
\beq\label{transition}
	U^{(21)}:\check\uD^{(2)}_{(0,2\pi-\alpha)} \to
	\check\uD^{(1)}_{(0,2\pi-\alpha)} ,\quad
	U^{(10)}:\check\uD^{(1)}_{(-\alpha,0)} \to
	\check\uD^{(0)}_{(-\alpha,0)},\quad
	U^{(02)}:\check\uD^{(0)}_{(0,\alpha)} \to
	\check\uD^{(2)}_{(-\alpha,0)}
\eeq
with $U^{(21)}(z)=z$, $U^{(10)}(z)=z$ and $U^{(02)}(z) = e^{-i\alpha}z$. It is clear that ${\cal M}_{\alpha,\ell}$ has excess angle $\alpha$ at the origin: it is a manifold with a conical singularity at the origin. Point II above may then be expressed as follows. Consider a product of local fields $\prod_j \Or_j(x_j)$ for disjoint coordinates $x_j\in\ell\uD\setminus\{0\}$ lying in the disk of radius $\ell$ minus the origin. Then,
\beq\label{tman}
	\frc{\bra \cV_\alpha(0)\,\prod_j \Or_j(x_j)\ket_{\ell\uD}}{\bra \cV_\alpha(0)\ket_{\ell\uD}} = \lt\bra \prod_j \Or_j(U(x_j)) \rt\ket_{{\cal M}_{\alpha,\ell}}
\eeq
for any $\Or_j(x_j)$, where $U:\ell\uD\setminus\{0\}\to {\cal M}_{\alpha,\ell}$ maps $\ell\uD\setminus[0,\ell)$ into $\check\uD^{(1)}$ and maps $(0,\ell)$ into $\check\uD^{(0)}$, in both cases as $U(z) = z$. Insertions of local fields in other regions of ${\cal M}_{\alpha,\ell}$ are obtained by continuation in the positions $x_j$.

\medskip
\noindent {\bf Remark.} When the angle $\alpha$ is an integer multiple of $2\pi$, it is possible to connect conical fields $\mathcal{V}_\alpha$ with ordinary twist fields, in a replica model, associated with special elements of the permutation group. Consider for instance the two-point function $\bra\mathcal{V}_\alpha(x) \mathcal{V}_{\alpha'}(0)\ket$ for $\alpha = 2\pi(n-1)$ and $\alpha' = 2\pi(n'-1)$ with $n,n'\in \N$. Consider a replica model composed of $m=n+n'-1$ copies of the original model, and the twist fields $\mathcal{T}_{(1\cdots n)}$ and $\mathcal{T}_{(n\cdots m)}$ associated to the permutation elements $(1\cdots n)$ and $(n\cdots m)$ that, respectively, cyclically permute the copies $1,\ldots,n$, and the copies $n,\ldots,m$. Then $\bra\mathcal{V}_\alpha(x) \mathcal{V}_{\alpha'}(0)\ket = \bra\mathcal{T}_{(1\cdots n)}(x) \mathcal{T}_{(n\cdots m)}(0)\ket$. In effect, the copy number $n$ is identified with the original plane, and the extra copies below and above $n$ with the extra space introduced by the conical singularities. Such constructions however only work for excess angles that are integer multiples of $2\pi$, although a natural analytic continuation in $n$ and $m$ of $\bra\mathcal{T}_{(1\cdots n)}(x) \mathcal{T}_{(n\cdots m)}(0)\ket$ will give the correct continuation to other values of $\alpha$ and $\alpha'$.

\section{Form Factor Approach to Conical Twist Fields}
In massive QFT an alternative way to define the conical twist fields is to fully characterise their matrix elements.
We therefore turn to the description of 1+1-dimensional QFT on Minkowski space-time in terms
of its Hilbert space of asymptotic relativistic particles. We will verify that (in the UV or short-distance limit) the correct OPEs \eqref{2}, and in particular \eqref{4}, are recovered, thus lending support to the fact that the field whose form factors we describe here is indeed the conical twist field introduced above.

In the context of $1+1$-dimensional QFT, form factors are defined
as tensor valued functions representing matrix elements of some
(local) operator $\mathcal{O}(x)$ located at the origin $x=0$
between a multi-particle {\em{in}}-state and the vacuum:
\begin{equation}
F_{k}^{\mathcal{O}|\mu _{1}\ldots \mu _{k}}(\theta _{1},\ldots
,\theta _{k}):=\left\langle
0|\mathcal{O}(0)|\theta_1,\ldots,\theta_k\right\rangle_{\mu_1,\ldots,\mu_k}^{\text{in}}
~.\label{ff}
\end{equation}
Here $|\theta_1,\ldots,\theta_k\rangle_{\mu_1,\ldots,\mu_k}^{\text{in}}$ represent
the physical ``in'' asymptotic states of massive QFT and $\langle 0|$ is the vacuum state. 
Multi-particle states carry indices $\mu_i$, which are quantum numbers characterizing the
various particle species, and depend on the real parameters
$\theta_i$, which are the associated rapidities. The rapidities characterize the energy and momenta of particles in 1+1 dimensions through the well-known relations 
$E_{\mu}=m_{\mu}\cosh \theta$ and $p_\mu=m_{\mu} \sinh\theta$, where $m_\mu$ is the mass of particle $\mu$. The form factors are
defined for all rapidities by analytically continuing from some
ordering of the rapidities; a fixed ordering provides a complete
basis of states, for instance $\theta_1>\theta_2>\cdots>\theta_k$ is the standard choice to describe {\it{in}}-states. 

In 1+1-dimensional integrable QFT there is a well-known approach known as the {\it form factor programme} which provides a systematic means to (a priori) compute all matrix elements (\ref{ff}) for any local field $\mathcal{O}$ \cite{KW,smirnovbook}. The ``standard" form factor programme generally relies on locality of operators. Nonetheless, we will now see that we can also develop such a programme for the conical twist fields by slightly altering and adapting the form factor axioms found in \cite{KW, smirnovbook}.

\begin{figure}[h!]
\begin{center}
\includegraphics[scale=0.3]{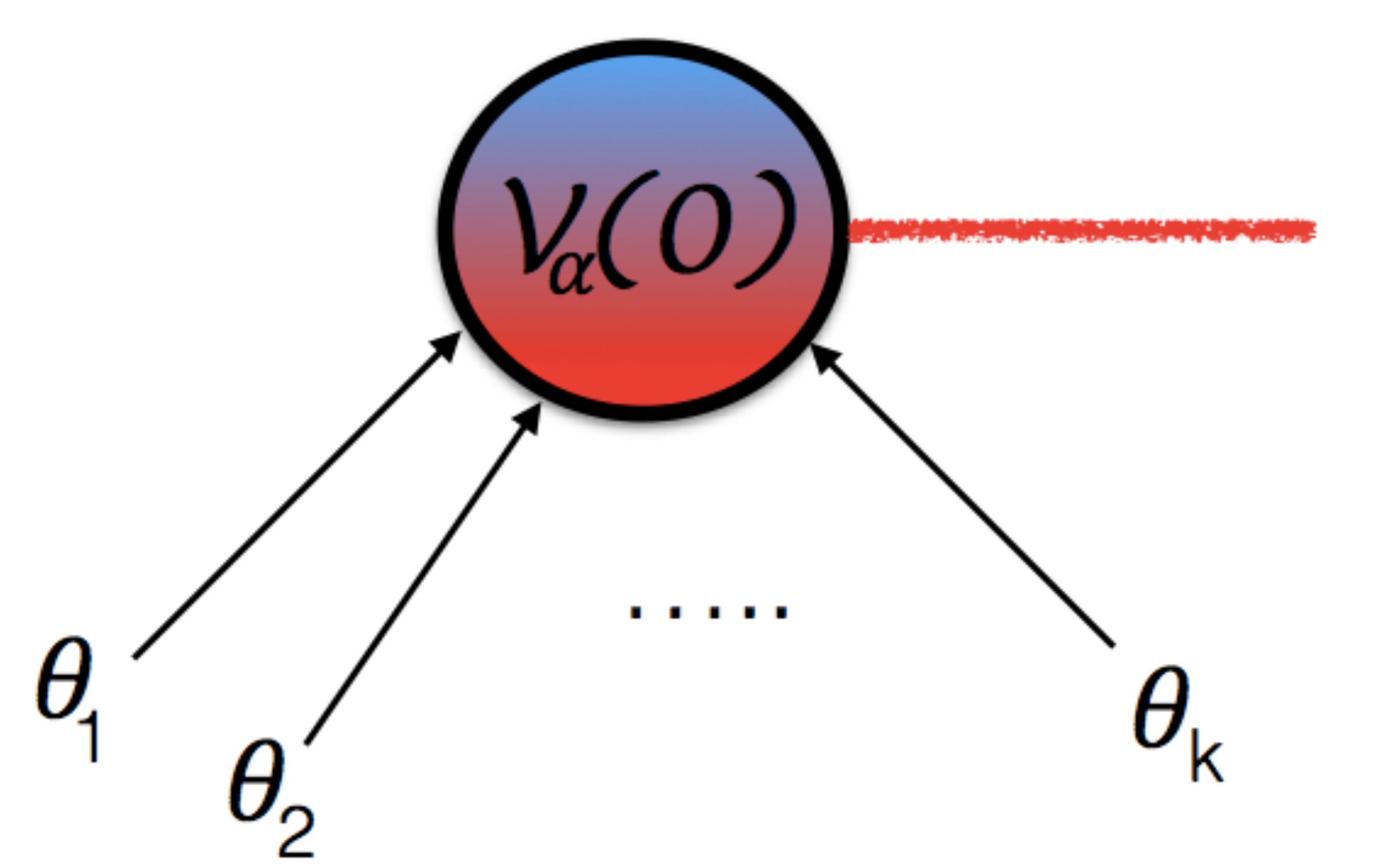} 
\caption{A pictorial representation of the conical twist field $k$-particle form factor. The red horizontal line represents the branch cut induced by the conical singularity.}
\label{ffV}
\end{center}
\end{figure}

Let us denote by
\begin{equation}
F_{k}^{\mathcal{V}_\alpha|\mu _{1}\ldots \mu _{k}}(\theta _{1},\ldots
,\theta _{k}):=\left\langle
0|\mathcal{V}_\alpha(0)|\theta_1,\ldots,\theta_k\right\rangle_{\mu_1,\ldots,\mu_k}^{\text{in}}
\end{equation}
the form factors of the conical twist field with excess angle $\alpha$. We may represent these functions pictorially as shown in Fig.~\ref{ffV}.
Because of relativistic invariance and spinlessness of the conical twist fields, these form factors must only depend on the rapidity differences; 
in the two-particle case, they then become functions of only one variable, $\theta=\theta_1-\theta_2$ and so we will write
$F_2^{\cV_\alpha|\mu_1\mu_2}(\theta)$ for the two-particle form factor. 

The form factor programme is a Riemann-Hilbert problem for the form factors (\ref{ffV}), that is a set of consistency equations which establish their monodromy properties and singularity structure. It was shown in \cite{entropy} that 
the standard form factor equations \cite{KW,smirnovbook} can be modified to encompass (local) branch point twist fields  defined on replica theories. Note that the form factor equations derived in \cite{entropy} are in fact very similar to equations found previously in \cite{nieder}, although in the latter a different motivation (i.e.~describing the Unruh effect) was taken, with entirely different  interpretation of the solutions.

Before embarking into the derivation of generic form factor equations for conical twist fields we should also mention that various proposals already exist in the literature which for the most part deal with specific amplitudes (that is, particular choices of $\alpha$). In \cite{BSV2} a set of consistency equations for pentagon amplitudes were proposed, which appear very similar to our equations in the next subsection for the particular choice $\alpha=\frac{\pi}{2}$. 
The equations for octagon amplitudes (or $\alpha=2\pi$) have been recently proposed in \cite{octagon} building on previous work by the same authors \cite{general}, even though the equivalent of our equation (\ref{kre}) was not written in \cite{octagon} or \cite{BSV2}.  Many works also exist where expressions for particular amplitudes are given in the form of series expansions which are clearly reminiscent of form factor spectral decompositions such as those discussed in section 5. For instance in \cite{BDM, didi1,didi2} such expansions are obtained from an integrable lattice structure whereas in \cite{DF1,DF2} a different approach is taken. 

\subsection{The conical field form factor equations}

\begin{figure}[h!]
\begin{center}
\includegraphics[scale=0.3]{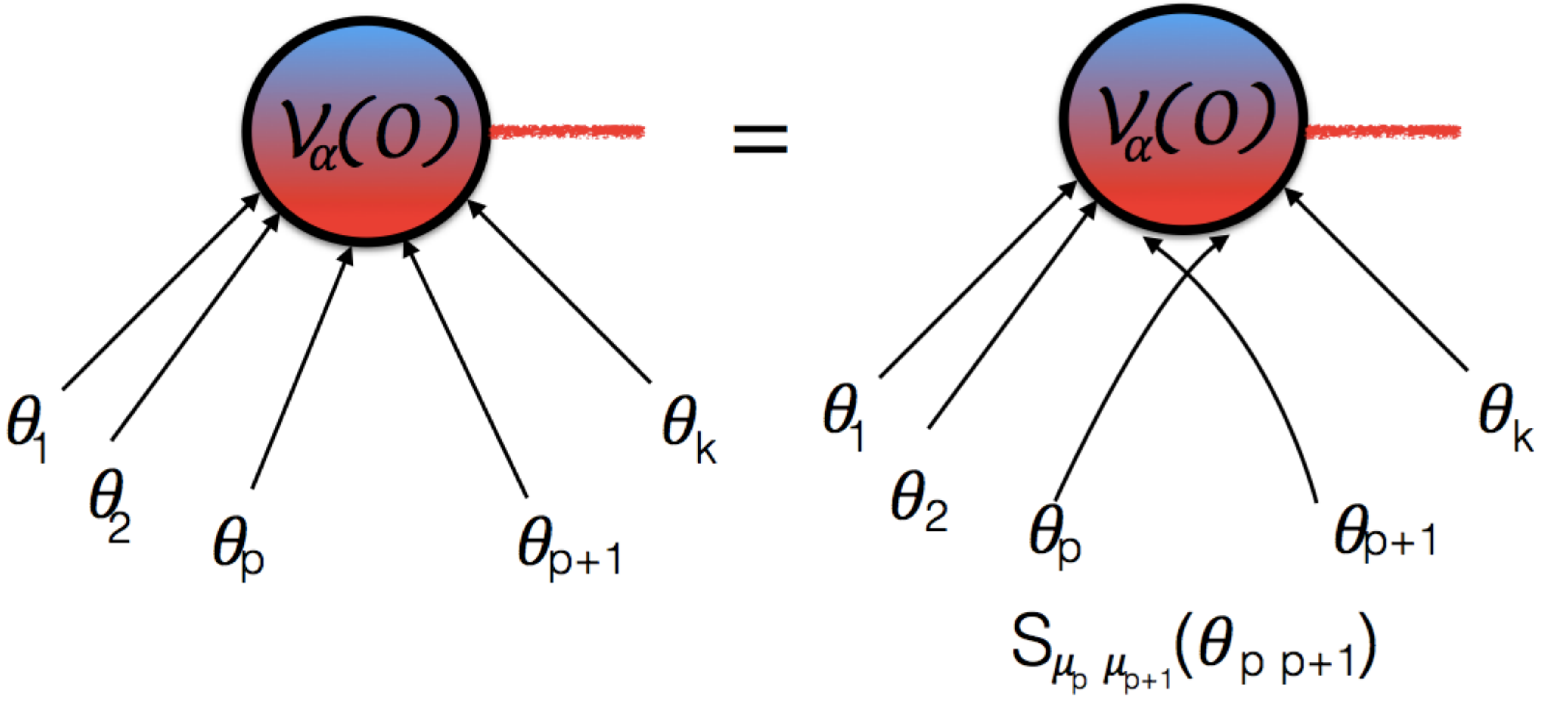} 
\caption{A Pictorial Representation of Watson's Equation (\ref{w1}). }
\label{f2}
\end{center}
\end{figure}

\begin{figure}[h!]
\begin{center}
\includegraphics[scale=0.3]{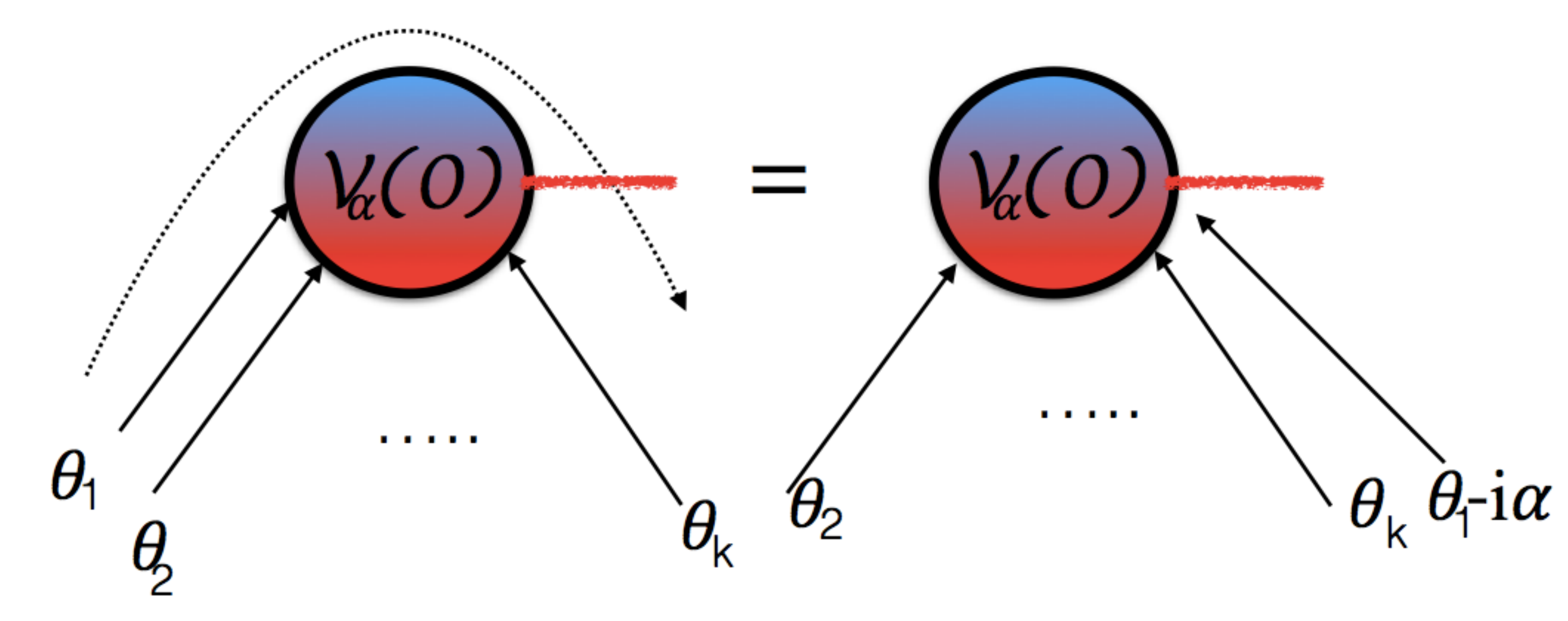}
\caption{A Pictorial Representation of Watson's Equation (\ref{w2}). }
\label{f3}
\end{center}
\end{figure}

\begin{figure}[h!]
\begin{center}
\includegraphics[scale=0.3]{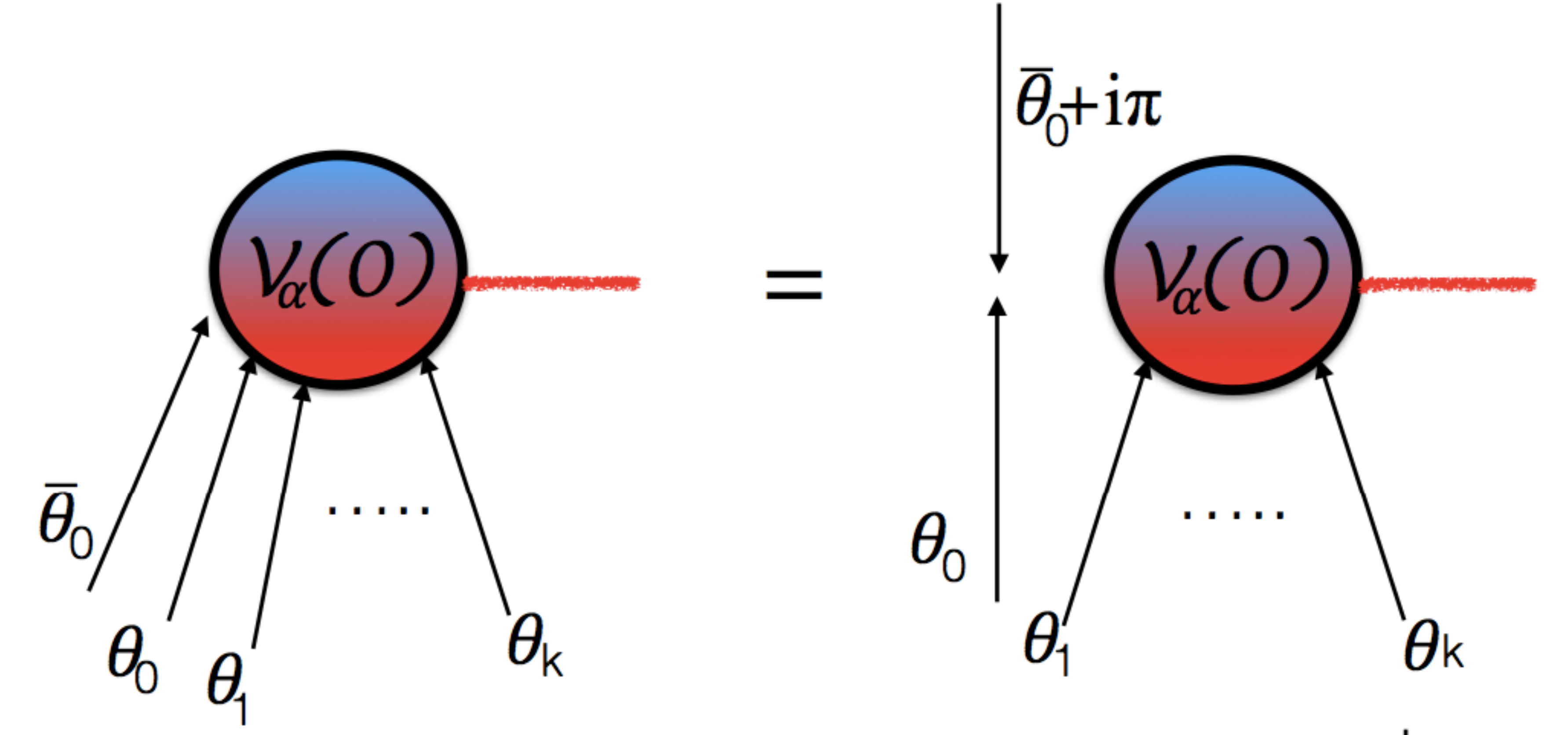} 
\caption{A Pictorial Representation of the kinematic residue equation (\ref{3}). }
\label{f4}
\end{center}
\end{figure}

\begin{figure}[h!]
\begin{center}
\includegraphics[scale=0.3]{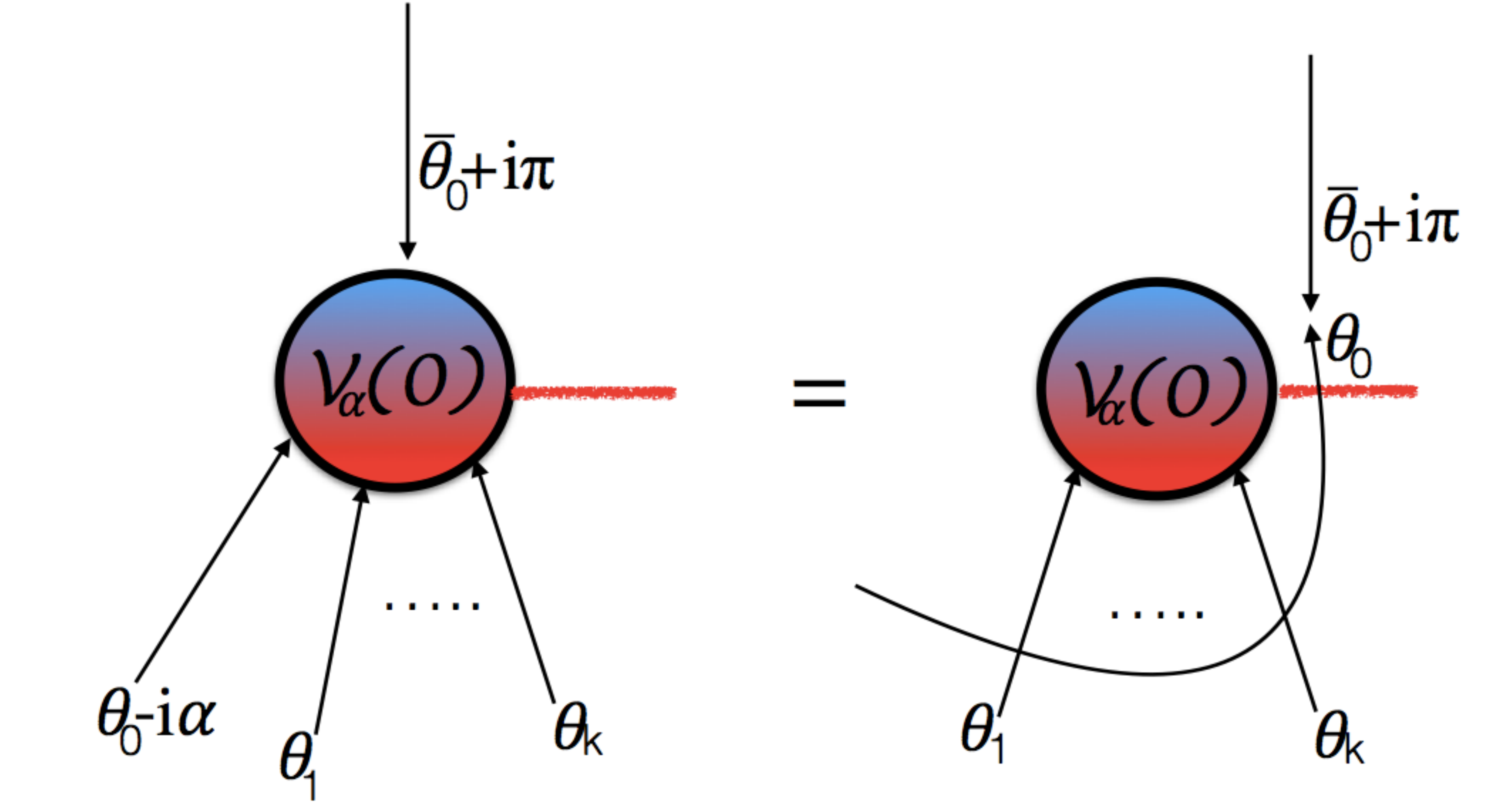} 
\caption{A Pictorial Representation of the kinematic residue equation (\ref{kre}). }
\label{f5}
\end{center}
\end{figure}

 It is intuitively not too hard to see how the equations from branch point twist fields may be adapted to the non-local conical twist field defined earlier. Let the two-particle scattering matrix between particles $\mu_1, \mu_2$ be $S_{\mu_1\mu_2}(\theta)$ (we assume for simplicity that there is no backscattering) and let  $\bar{\mu}$ represent the anti-particle of $\mu$. We propose the
following conical twist field form factor equations and represent them pictorially in Figs.~\ref{f2}, \ref{f3}, \ref{f4} and \ref{f5}:
 \beq
 F_{k}^{\mathcal{V}_\alpha|\ldots \mu_i  \mu_{i+1} \ldots }(\ldots,\theta_i, \theta_{i+1}, \ldots) =
  S_{\mu_i \mu_{i+1}}(\theta_{i\,i+1})
  F_{k}^{\mathcal{V}_\alpha|\ldots \mu_{i+1}  \mu_{i} \ldots}(\ldots,\theta_{i+1}, \theta_i,  \ldots),
  \label{w1}
 \eeq
\beq
 F_{k}^{\mathcal{V}_\alpha|\mu_1 \mu_2 \ldots \mu_k}(\theta_1+2 \pi \ri, \ldots,
\theta_k) =
  F_{k}^{\mathcal{V}_\alpha| \mu_2 \ldots \mu_k {\mu}_1}(\theta_2, \ldots, \theta_{k},
  \theta_1-\ri\alpha),\label{w2}\\
\eeq
 \beq
\underset{\bar{\theta}_{0}={\theta}_{0}}{\operatorname{Res}}
 F_{k+2}^{\mathcal{V}_\alpha|\b{\mu} \mu  \mu_1 \ldots \mu_k}(\bar{\theta}_0+\ri\pi,{\theta}_{0}, \theta_1 \ldots, \theta_k)
  =
  \ri \,F_{k}^{\mathcal{V}_\alpha| \mu_1 \ldots \mu_k}(\theta_1, \ldots,\theta_k), \label{3}
 \eeq
 \beq
\underset{\bar{\theta}_{0}={\theta}_{0}}{\operatorname{Res}}
 F_{k+2}^{\mathcal{V}_\alpha | \b\mu {\mu } \mu_1 \ldots \mu_k}(\bar{\theta}_0+\ri\pi,{\theta}_{0}-\ri\alpha, \theta_1 \ldots, \theta_k)
  =-\ri\prod_{i=1}^{k} S_{{\mu}\mu_i}(\theta_{0i}-\ri\alpha)
  F_{k}^{\mathcal{V}_\alpha| \mu_1 \ldots \mu_k}(\theta_1, \ldots,\theta_k).\label{kre}
\eeq

It is worth pointing out that equation (\ref{kre}) is actually not independent from (\ref{3}) (this is also true for the two kinematic residue equations for the branch point twist field given in \cite{entropy}). It is easy to show that (\ref{kre}) becomes (\ref{3}) after repeated use of 
(\ref{w1}), (\ref{w2}) and certain basic properties of the scattering matrix such as unitarity $S_{\mu_1\mu_2}(\theta)S_{\mu_2\mu_1}(-\theta)=1$ and crossing $S_{\mu_1\mu_2}(\theta+\ri\pi)=S_{\mu_2 \bar{\mu}_1}(-\theta)$. However, both equations (\ref{kre}) and (\ref{3}) are physically important since they imply the presence of two (rather than one) kinematic poles, and this has a bearing in determining the analytic structure of the form factors. Further use of \eqref{w1}, \eqref{w2},\eqref{3} and \eqref{kre} does not generate more kinematic poles. As usual, there will an additional residue equation in the case when bound state poles are present. This will be  identical to the branch point twist field bound state pole equation that can be found in \cite{entropy} (equation (6.8)) when restricted to particles on the same copy.

Let us now consider Watson's equations (\ref{w1})-(\ref{w2}) for the two-particle form factors. As noted earlier, these are functions of the rapidity differences only so we can write
 \beq
 F_{2}^{\mathcal{V}_\alpha|\mu_1\mu_2 }(\theta_{12} ) =
  S_{\mu_1\mu_{2}}(\theta_{12})
  F_{2}^{\mathcal{V}_\alpha|\mu_{2}  \mu_{1}}(-\theta_{12} )
 \eeq
 with $\theta_{12}:=\theta_1-\theta_2$ and
 \beq
 F_{2}^{\mathcal{V}_\alpha|\mu_1 \mu_2 }(\theta_{12}+2 \pi \ri) =
  F_{2}^{\mathcal{V}_\alpha| \mu_2  {\mu}_1}(-\theta_{12}+\ri\alpha),
\eeq
or, combining both equations:
\beq 
F_{2}^{\mathcal{V}_\alpha|\mu_1\mu_2 }(\theta_{12}) =
  S_{\mu_1\mu_{2}}(\theta_{12})
  F_{2}^{\mathcal{V}_\alpha|\mu_{2}  \mu_{1}}(-\theta_{12})= F_{2}^{\mathcal{V}_\alpha| \mu_2  {\mu}_1}(-\theta_{12}+\ri(2\pi+\alpha)).
  \label{w12}
\eeq 
The kinematic residue equations on the other hand tell us that
\beq 
\underset{\bar{\theta}_{0}={\theta}_{0}}{\operatorname{Res}} F_{2}^{\mathcal{V}_\alpha|\b{\mu} \mu }(\bar{\theta}_0-{\theta}_{0}+\ri\pi)
  =
  \ri F_{0}^{\mathcal{V}_\alpha},
\eeq 
and 
\beq 
\underset{\bar{\theta}_{0}={\theta}_{0}}{\operatorname{Res}}F_{2}^{\mathcal{V}_\alpha | \b\mu {\mu }}(\bar{\theta}_0-{\theta}_{0}+\ri(\alpha+\pi))
  =-\ri F_{0}^{\mathcal{V}_\alpha},
\eeq
where $F_{0}^{\mathcal{V}_\alpha}=\langle \cV_\alpha \rangle$ is the vacuum expectation value of the conical twist field. It turns out that, under the identification \eqref{thetan}, these are exactly the same equations as satisfied by the two-particle form factor of a branch point twist field on an $n$-copy replica model, if the particles $\mu_1, \mu_2, \mu, \bar{\mu}$ are all set to lie in the same copy of the replica model. This means that the same general solution found in \cite{entropy} will also solve the equations above giving:
\beq 
F_2^{\cV_\alpha|\mu_1\mu_2}(\theta)=\frac{\langle \mathcal{V}_\alpha \rangle
\sin\frac{\pi}{n}}{2 n \sinh\left(\frac{\ri\pi-\theta}{2n}\right)\sinh\left(\frac{\ri\pi+\theta}{2n}\right)}
\frac{F_{\text{min}}^{\mathcal{V}_\alpha|\mu_1\mu_2}(\theta)}{F_{\text{min}}^{\mathcal{V}_\alpha|\mu_1\mu_2}(\ri \pi)},\label{full}
\eeq 
where $\alpha$ is related to $n$ as per \eqref{thetan}, and
where $F_{\text{min}}^{\mathcal{V}_\alpha|\mu_1\mu_2}(\theta)$ is a minimal form factor, that is, a solution to equation (\ref{w12}) which has no poles on the extended physical sheet $\mathrm{Im}(\theta)\in [0,2\pi+\alpha]$. Here we have expressed (\ref{full}) in terms of the variable $n$ as formulae are simpler. 
We may now test the new form factor equations and their solutions by investigating the (massive) correlators of the fields whose conformal OPEs were given in (\ref{2})-(\ref{4}). Here for simplicity and in order to illustrate the concepts in the clearest fashion possible, we will do this for free theories, for which also higher particle form factors can be obtained. We will show that the CFT scaling behaviour is exactly recovered at short-distances from a form factor expansion. 

\subsection{Free Theories}
The simplest models to consider are of course free theories, that is the free Majorana Fermion and free Klein-Gordon Boson theories with relativistic scattering matrix $S(\theta)=\mp 1$, respectively. 
In this case there is only one particle type (so we may drop the particle indices $\mu_i$ in the form factors). Both these theories have an internal $\mathbb{Z}_2$ symmetries which implies only even-particle form factors are non-vanishing. The two-particle form factor is given by (\ref{full}) with
\begin{equation}
F_{\text{min}}^{\mathcal{V}_\alpha}(\theta)=\left\{ \begin{array}{cl}
-i\sinh\frac{\theta}{2n}& \mathrm{for \,\, the \,\, free\,\, Fermion}\\
1 & \mathrm{for \,\, the \,\, free \,\, Boson.}\\
\end{array}\right. \quad \label{is}
\end{equation}
Clearly, we have
\beq 
F_2^{\mathcal{V}_\alpha}(\theta)=\left\{ \begin{array}{cl}
-F_2^{\mathcal{V}_\alpha}(-\theta) & \mathrm{for \,\, the \,\, free\,\, Fermion}\\
F_2^{\mathcal{V}_\alpha}(-\theta) & \mathrm{for \,\, the \,\, free \,\, Boson.}\\
\end{array}\right.  \quad \label{is2}
\eeq 
Due to the free nature of these theories, and in particular the fact that the conical twist field is an exponential of quadratic expressions in free fields, the $k$-particle form factors admit remarkably simple expressions. They can be expressed as
\beq 
F_k^{\mathcal{V}_\alpha}(\theta_1, \ldots, \theta_k)= \left\{ \begin{array}{cl}
\langle \cV_\alpha \rangle{\mathrm{Pf}}(K)& \mathrm{for \,\, the \,\, free\,\, Fermion}\\
\langle \cV_\alpha \rangle{\mathrm{Perm}}(K) & \mathrm{for \,\, the \,\, free \,\, Boson,}\\
\end{array}\right.
\label{paf}
\eeq 
where $\mathrm{Pf}$ is the {\it Pfaffian} and $\mathrm{Perm}$ is the {\it Permanent} of the matrix $K$ defined as
\beq 
K_{ij}=\frac{{F_{2}^{\mathcal{V}_\alpha}(\theta_i-\theta_j)}}{\langle \cV_\alpha \rangle}.\label{K}
\eeq 
Recall that the Pfaffian of an antisymmetric matrix $K$ is given by
\beq 
{\rm{Pf}}(K)=\sqrt{\det(K)}
\eeq 
whereas the Permanent is given by
\beq 
{\rm{Perm}}(K)=\sum_{\sigma \in {S_k}} \prod_{i=1}^k K_{i\, \sigma(i)},
\eeq 
where $S_k$ is the set of all permutations of $\mathbb{Z}_k=\{1,2,\ldots,k\}$. Thus the two-particle form factor (\ref{K}) is the building block of all higher particle form factors of the conical twist field $\mathcal{V}_{\alpha}$. Since we will be using the expression (\ref{K}) repeatedly in the following computations, we will from now on use the shorter notation:
\beq 
K_{ij}:= f(\theta_{ij}; \alpha),
\eeq
with $\alpha$ given in (\ref{thetan}) and $\theta_{ij}=\theta_j-\theta_j$.

\section{Cumulant Expansion}
In 1+1-dimensional QFT we have that the (normalized) logarithm of the two-point function of local fields $\mathcal{O}_1, \mathcal{O}_2$ admits an expansion of the form \cite{Smir, takacs,karo}
\beq 
\log\left(\frac{\bra \mathcal{O}_1(0)\mathcal{O}_2(\ell) \ket}{\bra \mathcal{O}_1 \ket \bra \mathcal{O}_2 \ket} \right)= \sum_{j=1}^\infty c^{12}_j(\ell), \label{cs}
\eeq 
with
\beqa 
c^{12}_j(\ell)&=&\frac{1}{j!(2\pi)^{j} } \sum_{\mu_1,\ldots, \mu_{j}=1}^N \int_{-\infty}^\infty d\theta_1 \cdots \int_{-\infty}^\infty d\theta_{j} \, \, {h}^{12|\mu_1\ldots \mu_{j}}_{j}(\theta_1,\cdots,\theta_{j}) e^{-m\ell \sum_{i=1}^{j}\cosh\theta_i},
\label{genc}
\eeqa 
where the functions $h_j^{12|\mu_1\ldots \mu_j}(\theta_1,\cdots, \theta_j)$ are given in terms of the form factors of the fields involved, $N$ is the number of particles in the spectrum and $\mu_i$ represent the particle's quantum numbers. For example:
\beqa 
  h_1^{12|\mu}(\theta)&=& \frac{F_1^{\mathcal{O}_1|\mu}(\theta) \Big(F_1^{\mathcal{O}_2^\dagger|\mu}(\theta)\Big)^*}{\bra \mathcal{O}_1 \ket \bra \mathcal{O}_2\rangle}\nonumber\\
 h_2^{12|\mu_1 \mu_2}(\theta_{1},\theta_2)&=&\frac{{F}^{\mathcal{O}_1|\mu_1 \mu_2}_2(\theta_{1},\theta_{2})\Big({F}^{\mathcal{O}_2^\dagger|\mu_1 \mu_2}_2(\theta_1,\theta_2)\Big)^*}{\bra \mathcal{O}_1 \ket \bra \mathcal{O}_2 \rangle}-h_1^{12|\mu_1}(\theta_1)h_1^{12|\mu_2}(\theta_2), \label{exh}
\eeqa 
and so on. Here we have used the generic property:
\beq 
{\,}_{\mu_j\cdots \mu_1}\bra \theta_j \ldots \theta_1|\mathcal{O}_2(0)|0\ket=\bra 0 |\mathcal{O}_2^\dagger(0)|\theta_1\ldots\theta_j\ket_{\mu_1\cdots \mu_j}^* =:F_j^{\mathcal{O}_2^\dagger|\mu_1\ldots \mu_j}(\theta_1,\ldots,\theta_j)^*. \label{conjugation}
\eeq 

The expansion (\ref{genc}) with (\ref{exh}) is usually referred to as the cumulant expansion of the two-point function (see e.g.~\cite{Smir,takacs,karo}) and it is particularly well suited for extracting the leading $\log \ell$ behaviour of the two-point functions for $m\ell \ll 1$ (where $m$ is a mass scale) provided that the cumulants ${h}^{12|\mu_1\ldots \mu_{j}}_{j}(\theta_1,\cdots,\theta_{j})$ satisfy certain asymptotic properties in the rapidities so that the integrals (\ref{genc}) are finite. For simplicity let us consider temporarily the case of a single particle specie, and operators $\mathcal{O}_1, \mathcal{O}_2$ that are spinless. If all form factors are known, one may extract the leading UV behaviour by employing the fact that relativistic invariance implies that all form factors depend only on rapidity differences. As a consequence, one of the rapidities in the integrals (\ref{genc}) may be integrated over, leading to 
\beqa 
c^{12}_j(\ell)&=&\frac{2}{j!(2\pi)^{j} } \int_{-\infty}^\infty d\theta_2 \cdots \int_{-\infty}^\infty d\theta_{j} \, \, {h}^{12}_{j}(0,\theta_2,\cdots,\theta_{j}) K_0(m\ell d_j),
\label{genc222}
\eeqa 
where $K_0(x)$ is a Bessel function and 
\beq 
d_j^2=\left(\sum_{p=2}^j \cosh \theta_p +1\right)^2-\left(\sum_{p=2}^j \sinh \theta_p\right)^2. \label{dj}
\eeq 
Provided the functions ${h}^{12}_{j}(0,\theta_2,\cdots,\theta_{j})$ vanish for large $\theta_k$s, we may, for $m\ell \ll 1$, expand the Bessel function as $K_0(m\ell d_j)=-\log\ell-\gamma+\log 2-\log (m d_j)+\cdots$ where $\gamma=0.5772157...$ is the Euler-Mascheroni constant. 
For $m\ell \ll 1$ we expect the behaviour
\beq 
\log\left(\frac{\bra \mathcal{O}_1(0)\mathcal{O}_2(\ell) \ket}{\bra \mathcal{O}_1 \ket \bra \mathcal{O}_2 \ket} \right)_{m\ell \ll 1}= -x_{12}\log\ell-
{\cal K}_{12}+ \cdots \label{o1o2}
\eeq 
where $x_{12}$ is a constant that captures the short-distance power law of the correlator (described by CFT) and ${\cal K}_{12}$ is a constant
generally related to the expectation values and conformal structure constants of $\mathcal{O}_1$ and $\mathcal{O}_2$.

Coming back to the general case with potentially many particle species, using the leading term in the Bessel function expansion in the series (\ref{cs}) one obtains
\beqa
x_{12}=\sum_{j=1}^\infty \frac{2}{j!(2\pi)^{j}} \sum_{\mu_1,\ldots, \mu_{j}=1}^N \int_{-\infty}^\infty d\theta_2 \cdots \int_{-\infty}^\infty d\theta_{j} \, {h}^{12|\mu_1\ldots \mu_{j}}_{j}(0,\theta_2,\cdots,\theta_{j}).\label{del}
\eeqa
A similar expression can be found for the constant ${\cal K}_{12}$ as shown in \cite{karo}.

We note that the cumulant expansion method was applied in the context of gluon scattering amplitudes/WL in \cite{DF1, DF4}. In this context the O(6) NLSM is considered, and yields double-logarithmic terms ($\log\log$) in (\ref{o1o2}). Similar double-logarithm terms are also found in the branch-point twist field two-point function of the free boson model \cite{davide,olivier}. Nevertheless, the leading logarithm is still present, and the constant $x_{12}$ is still related to the short-distance dominant power law.

\subsection{Cumulant Expansion for Conical Twist Fields in Free Theories}
We use the cumulant expansion above, and obtain
\beq 
\log\left(\frac{\langle \mathcal{V}_\alpha(0) \mathcal{V}_{\alpha'}(\ell)\rangle}{\bra \mathcal{V}_\alpha \ket \bra \mathcal{V}_{\alpha'} \ket } \right)= \sum_{j=1}^\infty c^{\alpha \alpha'}_{2j}(\ell)   \label{cs2}
\eeq 
with
\beqa 
c^{\alpha \alpha'}_{2j}(\ell)=\frac{1}{(2j)!(2\pi)^{2j} }  \int_{-\infty}^\infty d\theta_1 \cdots \int_{-\infty}^\infty d\theta_{2j} \, \, {h}^{\alpha \alpha'}_{2j}(\theta_1,\cdots,\theta_{2j}) e^{-m\ell \sum\limits_{i=1}^{2j}\cosh\theta_i}.
\label{genc22}
\eeqa 
Compared to (\ref{genc222}) we have dropped the $\mu_1,\ldots, \mu_j$ super-indices in $h_j$ as from now on we will work only with one particle type. We also sum over even particle numbers only, as for free theories all other twist field form factors are zero. 

The structure of correlators similar to (\ref{cs2}) has been studied in much detail in \cite{nexttonext} for the free Fermion and in \cite{davide} for the free Boson, in the context of measures of entanglement, for branch point twist fields $\TT$ and $\tilde{\TT}$. This structure is extremely simple due to the form of the higher particle form factors summarized in (\ref{paf}). This is a unique feature of free theories which will allow us to write  simple close formulae for the cumulant expansion (\ref{genc22}). In order to better understand this structure let us consider a particular example, namely the four-particle contribution to the expansion (\ref{genc22}). We need to compute 
\beqa
{h}^{\alpha \alpha'}_{4}(\theta_1,\cdots,\theta_{4})&=& F_4^{\mathcal{V}_\alpha}(\theta_1,\theta_2,\theta_3, \theta_4)F_4^{\mathcal{V}_{\alpha'}}(\theta_1,\theta_2,\theta_3, \theta_4)^*-\left|F_2^{\mathcal{V}_\alpha}(\theta_{12})\right|^2 \left|F_2^{\mathcal{V}_{\alpha'}}(\theta_{34}) \right|^2 \nonumber\\
&& - \left|F_2^{\mathcal{V}_\alpha}(\theta_{13})\right|^2 \left|F_2^{\mathcal{V}_{\alpha'}}(\theta_{24}) \right|^2-\left|F_2^{\mathcal{V}_\alpha}(\theta_{14})\right|^2 \left|F_2^{\mathcal{V}_{\alpha'}}(\theta_{23})\right|^2. 
\eeqa 
A standard property of free theories is that when the four-particle contribution above is written in terms of two-particle form factors (using (\ref{paf})) the three two-particle contributions above are identically cancelled and only six terms remain that can be written as
\beqa
{h}^{\alpha \alpha'}_{4}(\theta_1,\theta_2,\theta_3,\theta_{4})\stackrel{{\mathrm{int}}}= 6 f(\theta_{12};\alpha) f(\theta_{23};\alpha')^*f(\theta_{34};\alpha)f(\theta_{14};\alpha')^*,
\eeqa 
where the superscript ``int" means equality under integration in all rapidities (in other words, the six terms are all equal under relabelling of the rapidities). 
This property holds both for free Fermions and free Bosons and generalises to all functions $h^{\alpha \alpha'}(\theta_1,\cdots,\theta_{2j})$ with the prefactor $6$ above generalizing to $(2j-1)!$ so that we may write
\beqa 
{h}^{\alpha \alpha'}_{2j}(\theta_1,\cdots,\theta_{2j})&\stackrel{{\mathrm{int}}}=&(2j-1)! f(\theta_{12};\alpha) f(\theta_{1\,2j};\alpha')^*\prod_{k=1}^{j-1} f(\theta_{2k+1\,2k+2};\alpha) f(\theta_{2k\,2k+1};\alpha')^*\nonumber\\
&\stackrel{{\mathrm{int}}}=&(2j-1)! f(\theta_{12};\alpha) f(\theta_{2j\,1};\alpha')\prod_{k=1}^{j-1} f(\theta_{2k+1\,2k+2};\alpha) f(\theta_{2k+1\,2k};\alpha'),
\eeqa 
up to the identification $\theta_{2j+1}\equiv \theta_{1}$. In the second line we have used the fact that $f(\theta;\alpha)^*=-f(\theta;\alpha)=f(-\theta;\alpha)$ for free Fermions and $f(\theta;\alpha)^*=f(\theta;\alpha)=f(-\theta;\alpha)$ for free Bosons.  
We can therefore write
\beqa 
c^{ \alpha \alpha'}_{2j}(\ell)&=&\frac{1}{(2j)(2\pi)^{2j} } \int_{-\infty}^\infty d\theta_1 \cdots \int_{-\infty}^\infty d\theta_{2j} \,e^{-m\ell \sum_{i=1}^{2j}\cosh\theta_i}  f(\theta_{12};\alpha) f(\theta_{2j\,1};\alpha')\nonumber\\
&&\quad \quad \quad\quad \quad \quad\quad \quad \quad \quad \quad \quad\times \prod_{k=1}^{j-1} f(\theta_{2k+1\,2k+2};\alpha) f(\theta_{2k+1\,2k};\alpha'),
\label{realc}
\eeqa
which is valid both for free Fermions and Bosons.

\subsection{Exact Short-Distance Asymptotics of Correlation Functions}
As shown by equation (\ref{del}) and by comparing to the short-distance CFT predictions (\ref{2})-(\ref{4}) we may extract the short distance asymptotics 
\beq
\log\left(\frac{\langle \mathcal{V}_\alpha(0) \mathcal{V}_{\alpha'}(\ell)\rangle}{\bra \mathcal{V}_\alpha \ket \bra \mathcal{V}_{\alpha'} \ket } \right)_{m\ell \ll 1}=-x_{\alpha \alpha'} \log \ell + \log\left( \frac{\bra \mathcal{V}_{\alpha+\alpha'} \ket \, \mathcal{C}_{\alpha \alpha'}^{\alpha+\alpha'}}{\bra \mathcal{V}_\alpha \ket \bra \mathcal{V}_{\alpha'} \ket}\right). \label{log1}
\eeq 
Comparing to (\ref{2})-(\ref{4}) we expect that
\beq 
x_{\alpha \alpha'}=2\Delta_\alpha+2\Delta_{\alpha'}-2\Delta_{\alpha+\alpha'}.
\eeq 
On the other hand, combining the form factor expansions (\ref{realc}) with the general results (\ref{genc222}) and (\ref{del}) we obtain
\beqa 
x_{\alpha\alpha'}=\sum_{j=1}^\infty\frac{1}{j(2\pi)^{2j}} \int_{-\infty}^\infty d x_2 \cdots \int_{-\infty}^\infty d x_{2j} \,f(-\sum_{k=2}^{2j} x_{k};\alpha) f(x_{2j};\alpha') \prod_{k=1}^{j-1} f(x_{2k+1};\alpha) f(-x_{2k};\alpha')
\label{sumttt}
\eeqa 
where we changed variables to 
\beq 
x_k=\theta_{k,k+1} \quad \mathrm{for} \quad k=2,\ldots, 2j-1,
\eeq 
and $x_{2j}=\theta_{2j}$ we have, in particular that
\beq 
\theta_2=\sum_{k=2}^{2j} x_{k}.
\eeq 
The expressions above can be simplified further by factoring integrals depending on odd-labelled and even-labelled variables after introducing a new variable $y=\sum_{k=1}^{j} x_{2k}$ so that
\beqa 
x_{\alpha\alpha'}&=&\sum_{j=1}^\infty\frac{1}{j(2\pi)^{2j}} \int_{-\infty}^\infty d y \int_{-\infty}^\infty d x_2 \cdots \int_{-\infty}^\infty d x_{2j-1} \,f(-y-\sum_{k=1}^{j-1} x_{2k+1};\alpha) f(y-\sum_{k=1}^{j-1} x_{2k};\alpha') \nonumber
\\
&\times& \prod_{k=2}^j f(x_{2k-1};\alpha) f(-x_{2k};\alpha')=\pm \sum_{j=1}^\infty\frac{1}{j(2\pi)^{2j}} \int_{-\infty}^\infty dy  \,G_j(y;\alpha) G_j(y;\alpha'),
\label{sumttty}
\eeqa 
where the plus sign is for the free Boson, the minus sign corresponds to the free Fermion. Here
\beqa 
G_j(y;\alpha)=\int_{-\infty}^\infty d s_1 \ldots d s_{j-1} \, f(y+\sum_{k=1}^{j-1} s_{k};\alpha)\prod_{k=1}^{j-1} f(s_k;\alpha)= \int_{-\infty}^\infty d x \, G_{j-1}(y+x;\alpha) f(x;\alpha)
\eeqa 
for $j>1$ and $G_1(y;\alpha)=f(y;\alpha)$. 
The introduction of the functions $G_j(y;\alpha)$ together with the recursive relation above provides a powerful numerical recipe for the evaluation of the sum (\ref{sumttt}).  Indeed a similar recursive structure was exploited in \cite{nexttonext, davide} in order to find an even more efficient way to evaluate the integrals (\ref{sumttty}). This is based on using the relation
\beq 
\int_{-\infty}^\infty dy \, G_j(y;\alpha) e^{\ri y s} =  \hat{f}(s;\alpha)^j,
\eeq 
where $\hat{f}(s;\alpha)$ is the
Fourier-transformed two-particle form factor
\beq 
\hat{f}(s;\alpha)=\int_{-\infty}^\infty dy\, f(y; \alpha) e^{\ri y s}.
\eeq 
This in turn means that we can write
\beq 
G_j(y;\alpha) =\frac{1}{2\pi}\int_{-\infty}^\infty ds \hat{f}(s;\alpha)^j e^{-\ri s y},
\label{four}
\eeq 
so that the evaluation of every function $G_j(y;\alpha)$ is reduced to the computation of a single integral. Furthermore, once we have written $G_j(y;\alpha)$ in the form (\ref{four}) the infinite sum (\ref{sumttty}) can be computed 
exactly to:
\beqa 
x_{\alpha \alpha'}&=&\mp \frac{1}{(2\pi)^2}\int_{-\infty}^\infty dy \int_{-\infty}^\infty ds_1 \int_{-\infty}^\infty ds_2 \log\left(1-\frac{\hat{f}(s_1;\alpha) \hat{f}(s_2; \alpha')}{(2\pi)^2} \right) e^{-\ri y(s_1+ s_2)}\nonumber\\
&=& \mp \frac{1}{2\pi} \int_{\infty}^\infty ds \, \log\left(1-\frac{\hat{f}(s;\alpha) \hat{f}(-s; \alpha')}{(2\pi)^2} \right), \label{resum}
\eeqa 
where we used the definition $\delta(x-a)=\frac{1}{2\pi}\int_{-\infty}^\infty d y e^{-\ri y(x-a)}$ of the Dirac delta function. The minus sign corresponds to the free Boson theory and the plus sign to the free Fermion theory. 
All that is left now is to compute the function
$\hat{f}(s;\alpha)$. This is feasible for free theories. The function is not particularly simple but it is very easy to evaluate numerically in the context of the integral above. We have
\beqa 
\!\!\!\!\!\!\!\!\!\!\!\!\!\!\hat{f}(s;\alpha) &= & \frac{1}{ns+\ri}\left[{e^{-\frac{\ri\pi}{n}}}{\,}_2F_1(1,1-\ri ns,2-\ri ns;e^{-\frac{\ri\pi}{n}})-{e^{\frac{\ri\pi}{n}}}{\,}_2F_1(1,1-\ri ns,2-\ri sn;e^{\frac{\ri\pi}{n}})\right]\nonumber\\
&-&  \frac{1}{ns-\ri}\left[{e^{-\frac{\ri\pi}{n}}} {\,}_2F_1(1,1+\ri ns,2+\ri ns;e^{-\frac{\ri\pi}{n}})-{e^{\frac{\ri\pi}{n}}} {\,}_2F_1(1,1+\ri ns,2+\ri ns;e^{\frac{\ri\pi}{n}})\right],
\eeqa 
for the free Boson and 
\beqa
\hat{f}(s;\alpha)&=& \frac{\csc\frac{\pi}{2n}}{2\ri ns-3} \left[e^{-\frac{\ri\pi}{n}}{\,}_2F_1(1,\frac{3}{2}-\ri ns,\frac{5}{2}-\ri ns;e^{-\frac{\ri\pi}{n}})-{e^{\frac{\ri\pi}{n}}}{\,}_2F_1(1,\frac{3}{2}-\ri ns,\frac{5}{2}-isn;e^{\frac{\ri\pi}{n}})\right]\nonumber\\
&-&  \frac{\csc\frac{\pi}{2n}}{1+2nis} \left[e^{-\frac{\ri\pi}{n}}{\,}_2F_1(1,\frac{1}{2}+\ri ns,\frac{3}{2}+\ri ns;e^{-\frac{\ri\pi}{n}})-{e^{\frac{\ri\pi}{n}}}{\,}_2F_1(1,\frac{1}{2}+\ri ns,\frac{3}{2}+isn;e^{\frac{\ri\pi}{n}})\right]\nonumber\\
&+& \frac{\csc\frac{\pi}{2n}}{1-2nis} \left[e^{-\frac{\ri\pi}{n}}{\,}_2F_1(1,\frac{1}{2}-\ri ns,\frac{3}{2}-\ri ns;e^{-\frac{\ri\pi}{n}})-{e^{\frac{\ri\pi}{n}}}{\,}_2F_1(1,\frac{1}{2}-\ri ns,\frac{3}{2}-isn;e^{\frac{\ri\pi}{n}})\right]\nonumber\\
&+&  \frac{\csc\frac{\pi}{2n}}{2\ri ns+3} \left[e^{-\frac{\ri\pi}{n}}{\,}_2F_1(1,\frac{3}{2}+\ri ns,\frac{5}{2}+\ri ns;e^{-\frac{\ri\pi}{n}})-{e^{\frac{\ri\pi}{n}}}{\,}_2F_1(1,\frac{3}{2}+\ri ns,\frac{5}{2}+isn;e^{\frac{\ri\pi}{n}})\right],\n
\eeqa
for the free Fermion in term of Gauss' Hypergeometric function ${\,}_2F_1(a,b,c;z)$. 

Although the formulae are more involved \cite{karo,davide} it  is also possible to write down a form factor expansion of the constant term on the right hand side of (\ref{log1}). 
This would provide information about particular ratios of expectation values and structure constants. We will not consider this quantity here but it will be interesting to study it in the future and to understand its significance within the application of conical fields to the computation of scattering amplitudes of gluons in planar $\mathcal{N}=4$ SYM theory.

\section{Numerical Results}
In this section we present a comparison between the numerical evaluation of the formulae given in the previous section and the analytical formulae predicted from CFT. 
\subsection{The case $\alpha,\alpha' \geq 0$}
\begin{figure}[h!]
\label{fig6} 
\begin{center}
\includegraphics[scale=0.75]{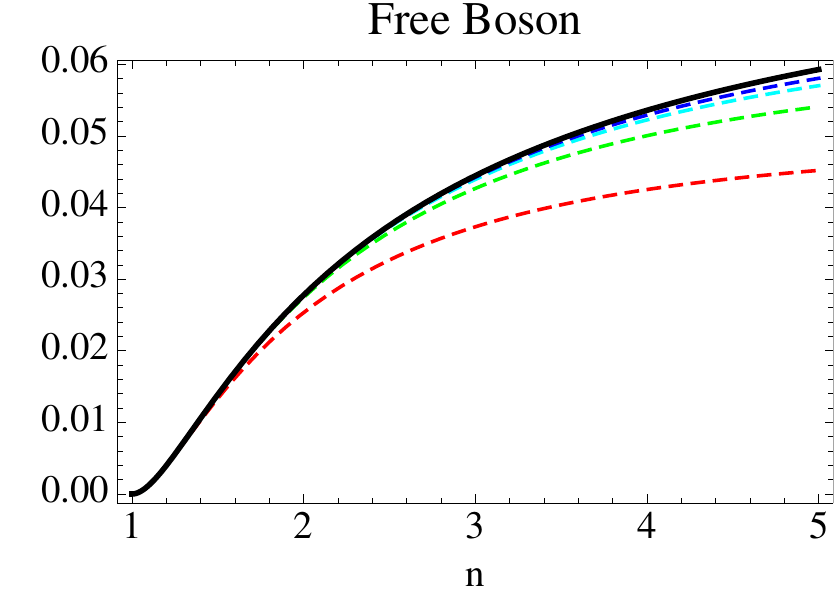} \,\,
\includegraphics[scale=0.77]{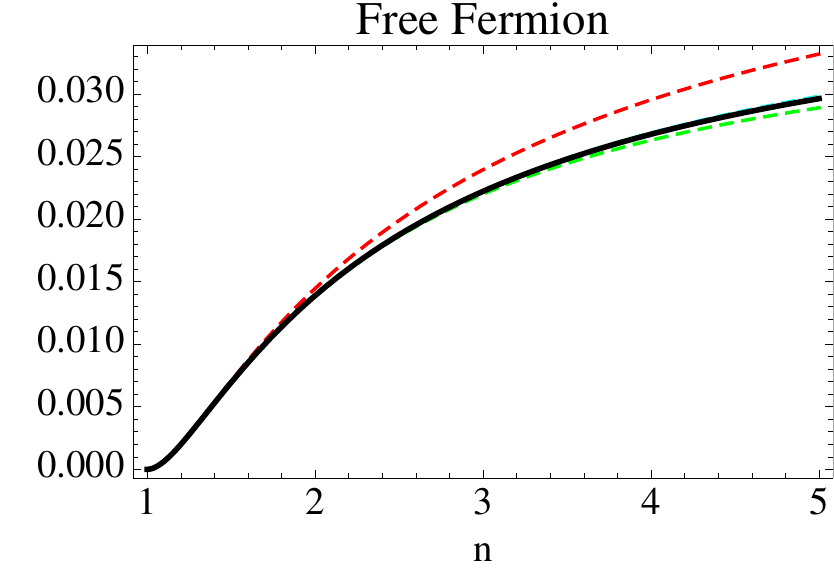}
\caption{The solid curves represent the function $x_{\alpha\alpha}=4\Delta_\alpha-2\Delta_{2\alpha}$ for the free Boson and free Fermion, respectively. They are evaluated  exactly from the formula (\ref{delta}) for different values of $n=\frac{\alpha}{2\pi}+1$. The dashed curves represent different approximations $\sum_{k=1}^{k_{\mathrm{max}}} x_{\alpha\alpha}^{(k)}$ of these values through the form factor expansion (\ref{sumttty}). The red dashed curve corresponds to $k_{\mathrm{max}}=1$, the green dashed curve is the value $k_{\mathrm{max}}=2$, the cyan dashed curve is the value $k_{\mathrm{max}}=3$ and the dashed blue curve is the value $k_{\mathrm{max}}=4$. In the free Boson case we observe very clear convergence to the predicted value. In the free Fermion convergence is even better, but the series $\sum_{k=1}^{k_{\mathrm{max}}} x_{\alpha\alpha}^{(k)}$ is alternating so that adding terms to the series alternatively overshoots and undershoots the exact value. }
\label{f6}
\end{center}
\end{figure}

\begin{figure}[h!]
\begin{center}
\includegraphics[scale=0.75]{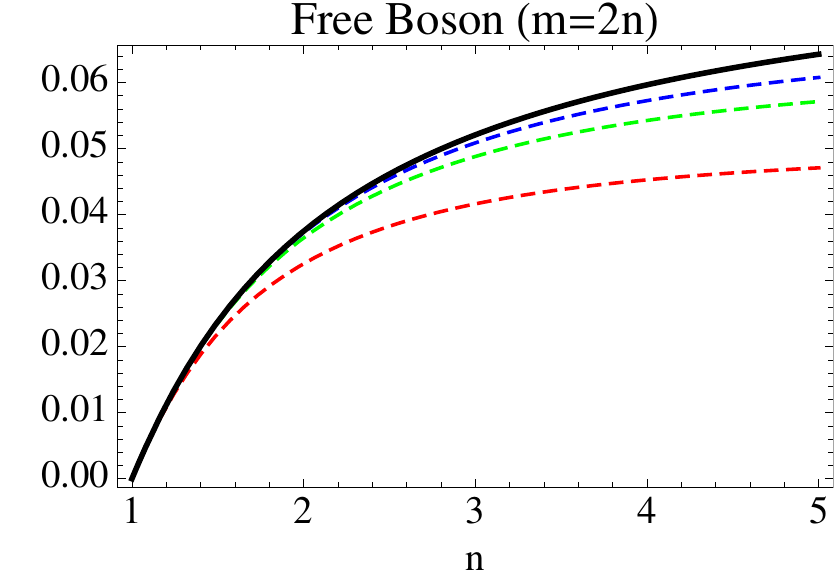} \,\,
\includegraphics[scale=0.77]{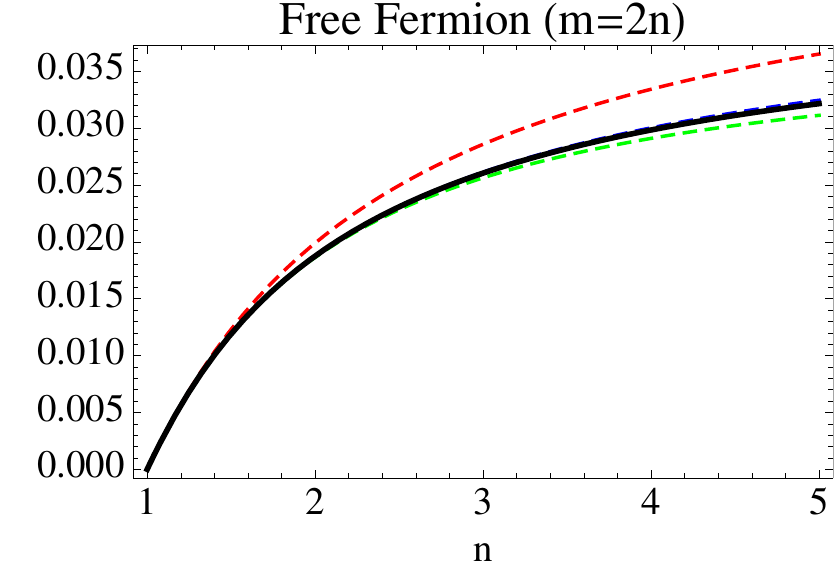}
\caption{The solid curves represent the function $x_{\alpha\alpha'}=2\Delta_\alpha+2\Delta_{\alpha'}-2\Delta_{\alpha+\alpha'}$ for $\alpha=2\pi(n-1)$ and $\alpha'=2\pi(2n-1)$ for the free Boson and free Fermion, respectively. They are evaluated  exactly from the formula (\ref{delta}) for different values of $n$. The dashed curves represent different approximations $\sum_{k=1}^{k_{\mathrm{max}}} x_{\alpha\alpha'}^{(k)}$ of these values through the form factor expansion (\ref{sumttty}). The red dashed curve corresponds to $k_{\mathrm{max}}=1$, the green dashed curve is the value $k_{\mathrm{max}}=2$, the blue dashed curve is the value $k_{\mathrm{max}}=3$. The convergence pattern is similar to that in Fig.~\ref{f6}. }
\label{f7}
\end{center}
\end{figure}

\begin{figure}[h!]
\begin{center}
\includegraphics[scale=0.75]{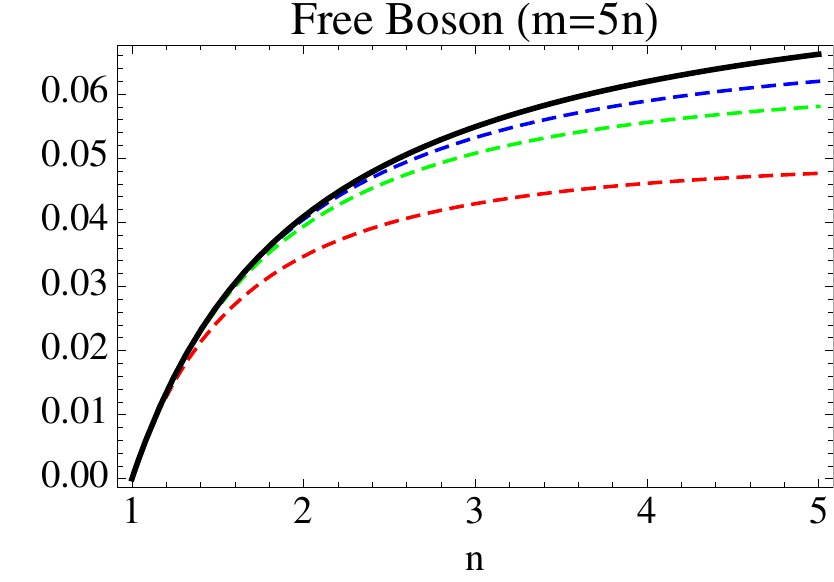} \,\,
\includegraphics[scale=0.77]{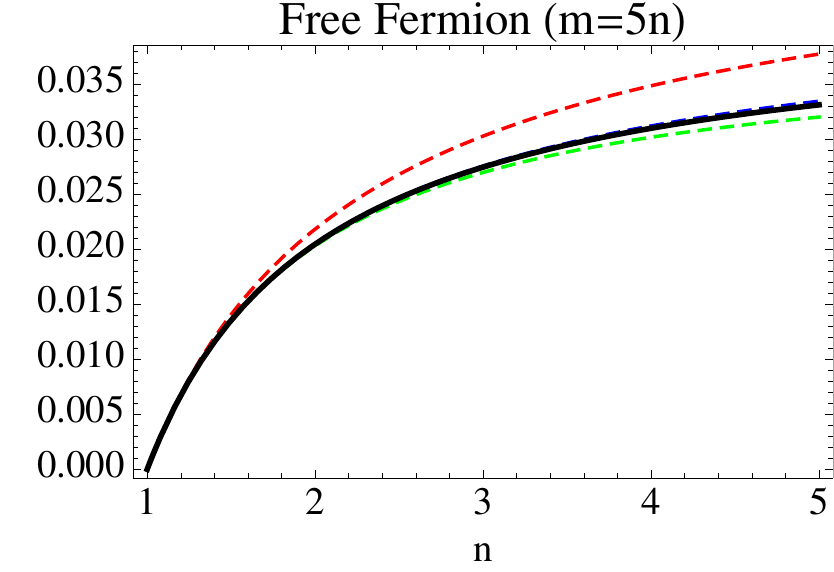}
\caption{The solid curves represent the function $x_{\alpha\alpha'}=2\Delta_\alpha+2\Delta_{\alpha'}-2\Delta_{\alpha+\alpha'}$ for $\alpha=2\pi(n-1)$ and $\alpha'=2\pi(5n-1)$ for the free Boson and free Fermion, respectively. They are evaluated  exactly from the formula (\ref{delta}) for different values of $n$. The dashed curves represent different approximations $\sum_{k=1}^{k_{\mathrm{max}}} x_{\alpha\alpha'}^{(k)}$ of these values through the form factor expansion (\ref{sumttty}). The red dashed curve corresponds to $k_{\mathrm{max}}=1$, the green dashed curve is the value $k_{\mathrm{max}}=2$, the blue dashed curve is the value $k_{\mathrm{max}}=3$. The convergence pattern is similar to that in Fig.~\ref{f6}, \ref{f7}. }
\label{f8}
\end{center}
\end{figure}

We may now study the sum (\ref{sumttt}) in more detail for free Fermions and Bosons and for particular choices of $\alpha$ and $\alpha'$. We will focus first on the case $\alpha, \alpha' \geq 0$ which corresponds to taking $n, n' \geq 1$.  The case $\alpha=\alpha'$ is of particular interest as this was the case considered in \cite{BSV}. Although the full re-summation \eqref{resum} can be done in free models, it is instructive, from a more general perspective, to write $x_{\alpha \alpha'}=\sum_{j=1}^\infty x_{\alpha\alpha'}^{(j)}$ and consider the first few contributions to $x_{\alpha \alpha}$ according to (\ref{sumttty}). For $j=1$ we have
\beq 
x_{\alpha \alpha}^{(1)}=\frac{1}{(2\pi)^{2}} \int_{-\infty}^\infty d x \,f(x;\alpha)f(-x;\alpha)=\left\{ \begin{array}{cc}
\frac{1}{2\pi n}\left(\frac{n-1}{n \sin\frac{\pi}{n}}-\frac{1}{ \pi}\right)& \mathrm{for \,\, the \,\, free\,\, Fermion}\\
\frac{1}{2 \pi n}\left(\frac{n-1}{n \tan \frac{\pi}{n}}+ \frac{1}{\pi}\right) & \mathrm{for \,\, the \,\, free \,\, Boson.}\\
\end{array}\right.
\eeq 
For $j=2$
\beqa 
x_{\alpha\alpha}^{(2)}=\pm\frac{1}{2(2\pi)^{4}} \int_{-\infty}^\infty d y \,\, G_2(y;\alpha)^2 \quad \mathrm{where} \quad 
G_2(y;\alpha)=\int_{-\infty}^\infty d x \, f(y+x; \alpha)f(x; \alpha),
\eeqa 
where again the plus sign corresponds to the free Boson and the minus sign corresponds to the free Fermion. It is possible to compute the function $G_2(y; \alpha)$ exactly:
\beqa 
G_2(y; \alpha)=\left\{ \begin{array}{cc}
\frac{2\left(4 \pi  (n-1) \tan \left(\frac{\pi }{2 n}\right) \cosh \left(\frac{y}{2 n}\right)-y
   \text{csch}\left(\frac{y}{2 n}\right) \left(\cosh \left(\frac{y}{n}\right)-2 \cos \left(\frac{\pi }{n}\right)+1\right)\right)}{n^2 \sec^2\left(\frac{\pi }{2n}\right) \left(\cos
   \left(\frac{2 \pi }{n}\right)-\cosh \left(\frac{y}{n}\right)\right)} & \mathrm{for \,\, the \,\, free\,\, Fermion}\\
\frac{4 \pi \left(1-n \right) \cot \left(\frac{\pi }{n}\right)-2{y \coth
   \left(\frac{y}{2n}\right)}}{n^2 \csc^2\left(\frac{\pi }{n}\right) \left(\cos \left(\frac{2 \pi }{n}\right)-\cosh \left(\frac{y}{n}\right)\right)} & \mathrm{for \,\, the \,\, free \,\, Boson.}\\
\end{array}\right.
\eeqa 
These analytical expressions can the be used to evaluate the integral for $x_{\alpha\alpha}^{(2)}$ numerically.  Analyzing other contributions, we obtain the numerical results shown in Fig.~\ref{f6}.
Similarly, we may consider other choices of $\alpha, \alpha'$. Some of these choices are shown in Figs.~\ref{f7}, \ref{f8}.
All the figures show that the form factor series is rapidly convergent. 

A stronger test can be carried out by employing directly the re-summed expression (\ref{resum}). In this case, the agreement is perfect over the whole range of  values of $n\geq 1$. Some examples are given below, see Fig. \ref{f9}.
\begin{figure}[h!]
\begin{center}
\includegraphics[scale=0.77]{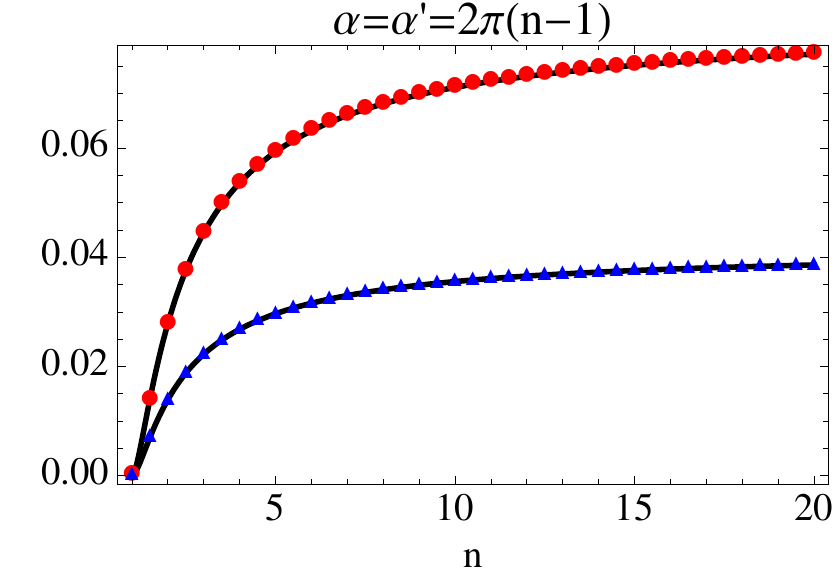} \,\,
\includegraphics[scale=0.77]{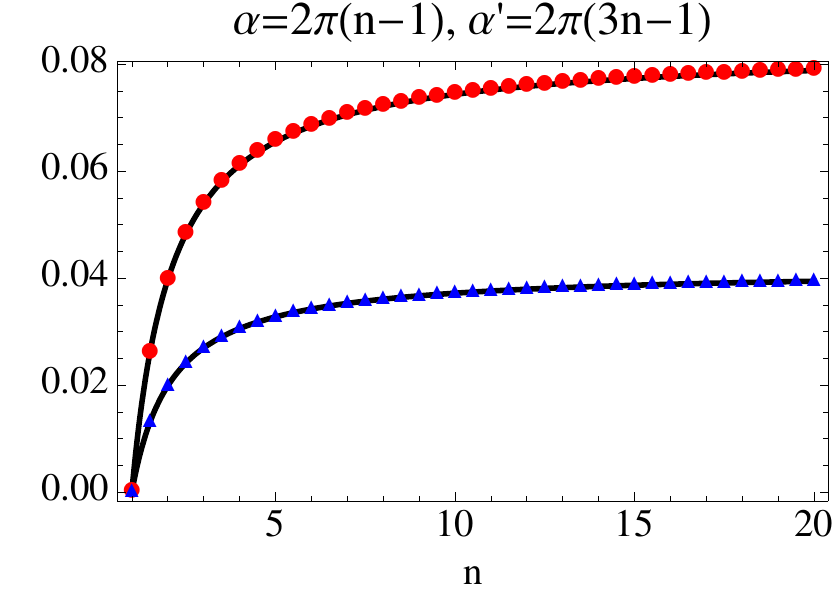}
\caption{The solid curves represent the function $x_{\alpha\alpha'}=2\Delta_\alpha+2\Delta_{\alpha'}-2\Delta_{\alpha+\alpha'}$ for different values of $\alpha$ and $\alpha'$ for the free Boson and free Fermion, respectively. 
The solid circles are the numerical outputs from integrating (\ref{resum}) for the free Boson and the solid triangles are the numerical values of (\ref{resum}) for the free Fermion. Agreement with the CFT prediction is perfect. }
\label{f9}
\end{center}
\end{figure}

\subsection{The case $\alpha,\alpha' < 0$}
\begin{figure}[h!]
\begin{center}
\includegraphics[scale=0.77]{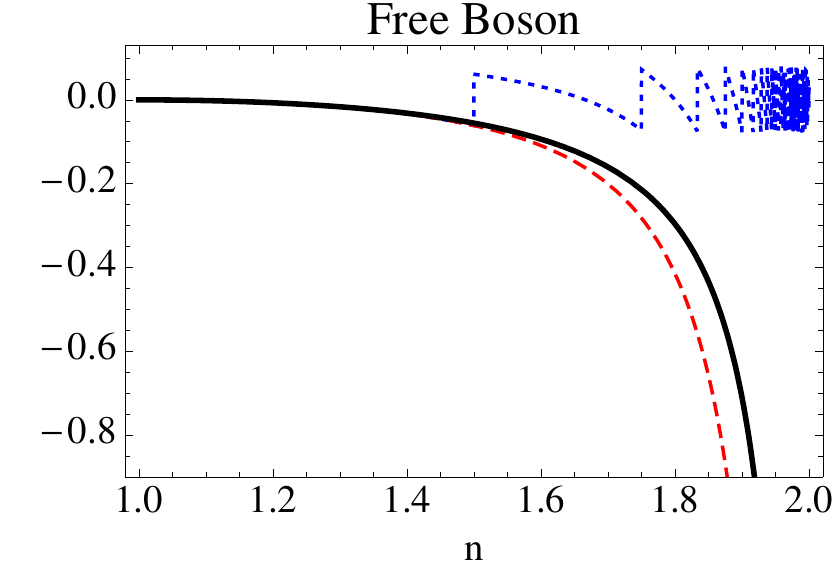} \,\,
\includegraphics[scale=0.77]{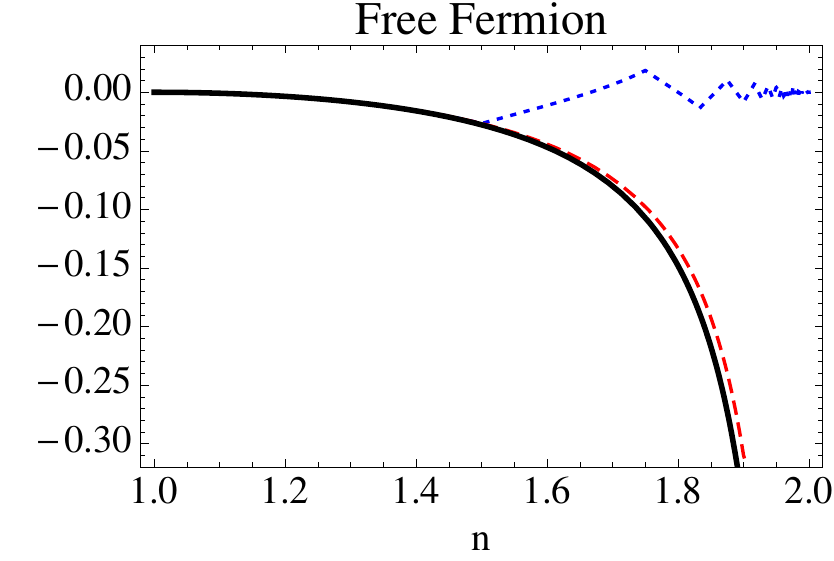}
\caption{The solid curves represent the function $x_{\alpha, -\alpha}=2\Delta_\alpha+2\Delta_{-\alpha}$ for different values of $\alpha$. 
The dotted lines are the result of integrating (\ref{x1}). There is good agreement for $n<1.5$ but above this value there are oscillations, far from the expected behaviour. The dashed curve is the ``corrected" integral (\ref{x12}), where the residues of all poles that cross the real line have been added to (\ref{x1}). }
\label{f10}
\end{center}
\end{figure}

In the examples above we have always considered positive excess angles. However we can also consider situations where the excess angles are negative.
Negative excess angles correspond to values of $n<1$ and this gives rises to some difficulties when trying to evaluate the form factor expansion (\ref{sumttty}). We can easily appreciate this if we try to evaluate the first contribution to $x_{\alpha, -\alpha}$. In this case, if $\alpha=2\pi(n-1)$ with $n\geq 1$ then $\alpha'=2\pi (1-n)<0$. Let us compute just the leading contribution to $x_{\alpha, -\alpha}$, given by
\beq 
x_{\alpha, -\alpha}^{(1)}=\frac{1}{(2\pi)^{2}} \int_{-\infty}^\infty d x \,f(x;\alpha)f(-x;-\alpha).
\label{x1}
\eeq 
This function can be evaluated numerically and the result can be compared to the CFT prediction $x_{\alpha, -\alpha}^{(1)}=2\Delta_{\alpha}+2\Delta_{-\alpha}=\frac{c (n-1)^2}{6 (n-2) n}$. Note that for $n=1$ we have $\alpha=0$ so there is no conical singularity and the conical twist fields are just the identity (so the power law is exactly 0). For $n=2$ on the other hand $-\alpha=-2\pi$ which makes no geometric sense, hence the singularity in the scaling dimensions. Let us just consider values $1\leq n<2$. The results are shown in Fig.~\ref{f10}.
The oscillations observed in Fig.~\ref{f10} reveal something about the structure of the integral (\ref{x1}). The picks occur for $n=2-\frac{1}{2m}$ with $m=1,2,\cdots$ These are the values of $n$ for which the function $f(-x;-\alpha)$ has a pole at $x=0$.  As we know, the functions $f(x;\alpha)$ have a kinematic pole structure with poles on the extended physical sheet at $x=\ri\pi$ and $x=\ri\pi (2n-1)=\ri(\pi+\alpha)$ and, more generally at all points of the form
 \beq
 x_k=\ri\pi(2nk \pm 1) \quad \mathrm{for}\quad  k=0, \pm 1, \ldots
 \eeq 
 For $n>1$ none of the poles above crosses the real line: $x_k=0$ for $n=\pm \frac{1}{2k}$ which is always less than 1.
  However, the situation is different for the function $f(-x;-\alpha)$. In this case the poles are
  \beq
 y_k=-\ri\pi(2k(2-n) \pm 1) \quad \mathrm{for} \quad k=0, \pm 1, \ldots
 \eeq 
and we have that $y_k=0$ if $n>1$ takes any of the values $n_k=2\pm \frac{1}{2k}$. 
At these precise values of $n$ the function $f(-x;\alpha)$ has a pole at $x=0$.  As $n$ varies from values smaller than $n_k$ to values above $n_k$ the corresponding pole $y_k$ 
moves from the upper (lower) half plane to the lower (upper) half plane. This means that the integral (\ref{x1}) must be corrected by adding the residues of all these poles. Similar issues arose and were studied in great detail in  \cite{nexttonext,davide} in the context of branch-point twist fields in free theories.
 This gives the expression
\beqa 
x_{\alpha, -\alpha}^{(1)}&=& \lim_{\epsilon \rightarrow 0}\frac{1}{(2\pi)^{2}} \int_{-\infty}^\infty d x \,f(x;\alpha)f(-x-i\epsilon;-\alpha)\nonumber\\
&& + \frac{1}{\pi} \sum_{k=1}^\infty (-1)^{s(k)} f(\ri\pi(2k (n-2)+1);\alpha)\Theta(2k(n-2)+1),
\label{x12}
\eeqa 
where $\Theta(x)$ is the Heavyside theta function which equals 1 for $x\geq 0$ and 0 otherwise, and $s(k)=k+1$ for the free Fermion and $s(k)=0$ for the free Boson. The effect of adding the residues of the poles that cross the real line is quite dramatic as can be see in Fig.~\ref{f10}. 
Similar, but more complicated corrections need to be added to every term in the expansion (\ref{sumttty}) to achieve agreement with the CFT prediction (\ref{4}). However, the leading term (\ref{x12}) already provides 
rather good agreement. 

\section{Conclusion and Outlook}
In this paper we have provided a rigorous definition of what we have termed conical twist fields. Our definition presents conical twist fields as twist fields associated with a space-time or external QFT symmetry, namely, in this precise case, the symmetry under rotation (by $\alpha$). Each conical twist field is therefore labelled by the rotation angle $\alpha$, and its insertion inside correlation functions introduces a conical singularity of excess angle $\alpha$. We propose that these are the fields whose correlators naturally arise in the study of gluon amplitudes/null polygonal Wilson loops in planar $\mathcal{N}=4$ SYM theory, at least for what concerns the scalar excitations \cite{BSV, DF1, DF4}. Yet, the present study gives us the suitable general view to imagine how this might extend to all the excitations (i.e.~to the whole gauge/string theory), raising the fascinating prospect of relating form factor series re-summations to thermodynamic Bethe ansatz computations, in the spirit of \cite{lodge}. 

We showed that the scaling dimensions \eqref{delta} of conical twist fields agree with those of the $n$-replica branch-point twist fields when $n$ is an integer and is related to $\alpha$ as in \eqref{thetan}. However, these two types of twist fields are different. Foremost, the branch point twist fields $\TT$, $\tilde{\TT}$ are defined in $n$-copy replica models and are associated to a generator of the $\Z_n$-symmetry of the replica model (that is, an internal symmetry of the replica QFT). The conical twist field does neither require $\alpha$ to be a multiple of $2\pi$ nor the model to consist of $n$ replicas. Furthermore, the hermiticity property \eqref{dag} and the OPE \eqref{2} differ from those of branch-point twist fields.

We have derived form factor equations for the conical twist fields and seen that
the solutions to these equations are identical to the form factors of the branch-point twist field \cite{entropy} when all particles lie on the same copy. 
As a consequence, the form factor series for the correlator 
$\bra\mathcal{V}_\alpha(0)\mathcal{V}_\alpha(\ell) \ket$ shares some features of the 
 form factor series for the vacuum two-point function $\langle \TT(0) \tilde{\TT}(\ell)\rangle_n$. However, importantly, the latter contains sums over replica indices, from 1 to $n$, as these parametrize the particles of the replica model. In contrast, that of conical twist fields does not involve such sums. This is exactly the feature observed in the series expansion obtained in \cite{BSV}. Given that two-point functions of conical and branch-point twist fields have so different form factor expansions, it is remarkable that both lead to the same scaling dimension, something which we have verified explicitly in free models.

It would be interesting to look at other types of non-local twist fields that may also have applications to the study of Wilson loops (gluon scattering amplitudes) in different situations. For instance, it is natural in theories with ``charged'' particles (as for instance gluons\footnote{Which are not properly charged, but change polarisation.} and fermions circulating in polygonal WLs) that the twist field induces, along with a rotation by $\alpha$, a particle charge conjugation\footnote{We thank Yunfeng Jiang, Ivan Kostov and Didina Serban for emphasizing this point to us.}. In fact, any other internal symmetry may be considered instead of charge conjugation, and it is easy to modify the form factor equations presented here to account for such ``hybrid" conical twist fields that combine rotation with internal symmetries, and study their correlators. 

Another set of important questions are those about correlation functions with fields present in the extra space afforded by a negative-curvature conical singularity. In particular, taking inspiration from works on null polygonal Wilson loops, one would expect that crossing symmetry gives access to various ``directions" of asymptotic particle states, for imaginary rapidity rotations by various angles between $\ri\pi$ and $\ri(\pi+\alpha)$. Augmenting conical fields with prescriptions on intermediate particles, it should also be possible to implement not only conical singularities, but also nontrivial branching: connections between branches are carried by particles lying within this extra rapidity space, in much the same way by which the sum over replicas in form factor expansions of branch-point twist field correlators reproduces branching of $n^{\rm th}$-root type.

Finally, the extended concept of twist fields introduced here can be applied to other space-time symmetries, such as scaling transformation (giving rise to ``spiral fields"). Studies of such fields would be very interesting.

\paragraph{Acknowledgments:} OCA and BD are grateful to Y. Jiang, I. Kostov and D. Serban for their interest in this work and for stimulating discussions during the summer school on ``Exact methods in low dimensional statistical physics" that took place at the Institute d'\'Etudes Scientifiques de Carg\`ese (Corsica) from July 25th to August 4th (2017).  They would like to thank the organizers of the school for the opportunity to carry out parts of this work in such a nice environment. BD thanks D. Bernard for comments, BD and OCA thank P. Vieira for a very interesting discussion and DF enjoyed fruitful discussions with A. Bonini, I. Kostov, S. Piscaglia and M. Rossi. OCA and BD are grateful to EPSRC for providing funding through the standard proposal ``Entanglement Measures, Twist Fields, and Partition Functions in Quantum Field Theory" under reference numbers EP/P006108/1 and EP/P006132/1. 
OCA and BD would also like to acknowledge financial support and hospitality by the Perimeter Institute for Theoretical Physics in Waterloo, Ontario (Canada) where part of this work was carried out. Generous
 financial support under an Emmy Noether Visiting Fellowship (OCA) and the PI Visitors Programme (BD) are gratefully acknowledged. OCA and BD also thank the Physics Department of the University of Bologna, INFN and Elisa Ercolessi for hospitality and financial support during an extended visit in November 2016 which led to this project's inception. 
DF was partially supported by the grants: GAST (INFN), UniTo-SanPaolo Nr TO-Call3-2012-0088, the ESF Network HoloGrav (09-RNP-092 (PESC)), the MPNS--COST Action MP1210 and the EC Network Gatis.
\appendix

\section{Main properties of conical twist fields.}\label{appcon}

In this appendix we provide QFT arguments showing the four properties of conical twist fields expressed in subsection \ref{ssectsingle}.

\subsection{Angular quantization}\label{sssectaq}

Angular quantization can be used efficiently in order to show properties I and IV.  We concentrate on the finite-volume setup, as the infinite-volume limit can be taken afterwards and does not affect the $\varep\to0$ limit.

In the angular quantization scheme \cite{freefield1,freefield2}, equal-time slices (``space'') are rays of length $\ell$ emanating from a center, which we choose to be the origin; and ``Euclidean time'' is the angle around it. In this scheme, the hamiltonian $K= \int_0^\ell \dd x\,\frak{R}^2(x)$ is proportional to the generator of rotations, (Euclidean) time is compact, and averages are evaluated by taking traces on the angular quantization space, $\bra\cdots\ket = \Tr\lt(e^{-2\pi K}\cdots\rt)/\Tr\lt(e^{-2\pi K}\rt)$. Here, on the right-hand side the ellipsis ``$\cdots$" are implicitly understood as the operators, in the angular-quantization representation, corresponding to the observables on the left-hand side.

The principal advantage of angular quantization for studying twist fields in general is that in the angular quantization where the center is its position, the twist field is represented as the operator implementing the {\em full} symmetry transformation to which it is associated. This is clear when the twist field is associated to a continuous symmetry with a Noether current $j^\mu$. Indeed, the twist field then has the form $\exp\big(\ri \alpha\int_0^\ell  \dd x\,j^0(x)\big)$; in angular quantization the quantity $\int_0^\ell  \dd x\,j^0(x)$ is the integral over all of space of a current density, hence it is the full conserved charge. The conserved charge can then be simultaneously diagonalized with the angular quantization hamiltonian $K$. The same principle holds for any twist field, independently from the existence of a Noether current. Hence the insertion of a twist field ${\cal T}_\sigma$, associated to a symmetry transformation $\sigma$, inside a correlation function, is formally evaluated, within angular quantization, as $\Tr \lt(e^{-2\pi K}{\cal U}_\sigma \cdots\rt)/\Tr\lt(e^{-2\pi K}\rt)$, where ${\cal U}_\sigma$ is the operator on the angular quantization space implementing the symmetry transformation.

The ratio of traces resulting from a twist field insertion needs renormalization. Following \cite{freefield1,freefield2}, it is convenient to make a small hole around the origin of radius $\varep$, and impose conformal boundary conditions on its boundary. The hamiltonian is now
\beq\label{Kham}
	K= \int_\varep^\ell \dd x\,{\frak R}^2(x).
\eeq
The $\varep$-regularization affects the angular quantization Hilbert space, so the traces depend on $\varep$. The limit $\varep\to0$ is singular, but for primary fields, the singularity is expected to be a power of $\varep$, and the power is expected to be the scaling dimension of the twist field.

Applying these ideas to the conical fields, correlation functions with a conical twist-field insertion are expected to be, in angular quantization, of the form
\beq\label{ang}
	\bra\cV_\alpha(0) \cdots\ket = \lim_{\varep\to0} \varep^{-2\Delta_{n}}\frc{\Tr_\varep\lt(e^{-2\pi n K}\cdots\rt)}{\Tr_\varep\lt(e^{-2\pi K}\rt)}
\eeq
with $\alpha$ related to $n$ as per \eqref{thetan}.

The proof of part I is obtained by showing that indeed \eqref{ang} is equivalent to \eqref{renang2}, and that the limit on $\varep$ in \eqref{ang} exists.

First, using \eqref{Kham}, we see that the integral in the exponential in \eqref{renang2} is exactly $-\alpha K$. This conserved charge is independent of the angular-quantization time, the angle $\phi$ around the origin. Therefore angle-splitted powers of $-\alpha K$, as per the prescription \eqref{regang}, are ordinary powers of this conserved charge, independent of angles. Therefore the limit in $\phi_k$ exists, and as a consequence, in angular quantization $\lt[e^{-\alpha\int_0^\infty \dd x \,{\frak R}^2(x)}\rt]_\varep$ is represented as $e^{-\alpha K}$, so that angular quantization indeed leads to \eqref{ang}.

We now show that the limit on $\varep$ exists in \eqref{ang}. Since the small-$\varep$ power law is obtained from short-distance behaviours, described by CFT, it is sufficient to specialize to the CFT case. Further, in QFT the short-distance behavior is not affected by the insertion of other fields at positions different from the origin (this is ``scale clustering'': microscopic divergencies factorize on local singularities). Therefore we may consider the one-point average of \eqref{renang2} on the disk of radius $\ell$, with the CFT expression of the rotation current \eqref{r0cft}. The parameter $\ell$ is now seen as an infra-red regulator, much smaller than all other scales in the initial expression (including the positions of the other fields and the correlation length). The boundary condition at $\ell$ can be chosen to be conformal.

The angular stress-energy tensor is obtained by performing the transformation to angular coordinates $\eta$ (``space'') and $\phi$ (imaginary ``time''), defined by $z=e^{\eta+i\phi}=:e^{\xi}$. The angular stress-energy tensor is then
\beq
	T^{\rm ang}(\eta,\phi) = T^{\rm ang}(\xi) = e^{2\xi} T(e^{\xi}) - \frc{c }{24},\quad
	\b T^{\rm ang}(\eta,\phi) = \b T^{\rm ang}(\b\xi) = e^{2\b\xi} \b T(e^{\b\xi}) - \frc{c }{24}.	
\eeq
The angular-quantization space variable takes values on the finite interval $\eta\in[\log\varep,\log \ell]$. On the boundaries, we impose conformal boundary conditions $T^{\rm ang}(\log\varep,\phi) = \b T^{\rm ang}(\log \varep,\phi)$, $T^{\rm ang}(\log \ell,\phi) = \b T^{\rm ang}(\log  \ell,\phi)$ for all $\phi\in[0,2\pi)$. The angular Hamiltonian is then
\beq\label{tang}
	K =
	-\frc1{2\pi}\int_\varep^\ell \dd x \lt(xT(x) + x\b T(x) - \frc{c}{6 x} \rt) =
	-\frc1{2\pi}\int_{\log \varep}^{\log \ell} \dd \eta \lt(T^{\rm ang}(\eta)
	+ \b T^{\rm ang}(\eta)\rt)
	+\frc{c}{24\pi} \log(\ell/\varep),
\eeq
and it is apparent that it is conserved thanks to the boundary conditions. Hence we have
\beq\label{dk}
	\frc{\p}{\p n}\log \Tr_\varep\lt(e^{-2\pi nK}\rt)
	= \int_{\log\varep}^{\log\ell} \dd\eta\,\bra T^{\rm ang}(\eta)
	+ \b T^{\rm ang}(\eta)\ket_{\varep,n}
		-\frc{c}{12} \log(\ell/\varep)
\eeq
where the average is defined by
\beq
	\bra\cdots\ket_{\varep,n} :=
	\frc{\Tr_\varep\lt(e^{-2\pi n K}\cdots
	\rt)}{\Tr_\varep\lt(e^{-2\pi n K}\rt)}.
\eeq

Clearly, the angular stress-energy tensor satisfies the same singular OPE's as the original one, for instance
\beq\label{TTxi}
	T^{\rm ang}(\xi)T^{\rm ang}(0)=
	\frc{c}{2\xi^4} + \frc{2 T^{\rm ang}(0)}{\xi^2}
	+ \frc{\p T^{\rm ang}(0)}{\xi} +O(1).
\eeq
This implies the expected commutation relations $[K,T(\eta,\phi)]=\p_\phi T(\eta,\phi)$  and $[K,\b T(\eta,\phi)]=\p_\phi \b T(\eta,\phi)$. Hence the average $\bra\cdots\ket_{\varep,n}$ is a CFT average on a finite cylinder of length $\log(\ell/\varep)$ and of circumference $2\pi n$ \footnote{This can be explicitly shown as follows. Consider two-point functions. Using the cyclic property of the trace, they are periodic in the imaginary direction with period $2\pi n$. The normalization of the fields is fixed by \eqref{TTxi}, and along with the boundary conditions, this is a Riemann-Hilbert problem with a unique solution. One-point functions on the cylinder are defined by the $1/\xi^2$ residue of the two-point function, and higher-point functions are analyzed similarly.}. This is mapped to the annulus with inner radius $\varep^{1/n}$ and outer radius $\ell^{1/n}$ via the map $\xi\mapsto y = e^{\xi/n}$, giving
\beq\label{ft}
	T^{\rm ang}(\xi) = e^{2\xi/n} T^{\rm an}(e^{\xi/n}) - \frc{c}{24n^2},\quad
	\b T^{\rm ang}(\b \xi) = e^{2\b \xi/n} \b T^{\rm an}(e^{\b \xi/n}) - \frc{c}{24n^2}.
\eeq
The stress-energy tensor has zero average on the annulus by rotation invariance, whereby we obtain
\beq
	\bra T^{\rm ang}(\eta)+ \b T^{\rm ang}(\eta) \ket_{\varep,n}
	= -\frc{c}{12n^2}.
\eeq
This leads to
\beq
	\frc{\p}{\p n} \log\Tr_\varep\lt(e^{-2\pi K}\rt)
	=- \frc{c}{12} \log(\ell/\varep)\,\lt(1+\frc1{n^2}\rt).
\eeq
Integrating,
\beq
	\Tr_\varep\lt(e^{-2\pi n K}\rt) = A\lt(\frc{\ell}{\varep}\rt)^{-\frc{c}{12}\lt(n-\frc1n\rt)}
\eeq
where $A$ is independent of $n$ (but may depend on $\ell$ and $\varep$). Hence, we have found \eqref{norm} and shown that the limit on $\varep$ in \eqref{ang} exists.

\subsection{Manifold reconstruction}\label{sssectman}

Using the twist properties of $\cV_\alpha$, one may explicitly reconstruct the manifold ${\cal M}_{\alpha,\ell}$ with an excess angle of $\alpha$ at the origin of the disk of radius $\ell$, showing property II. We give the proof of property II in the generic case of a disk $\ell\uD$, with rotation-invariant boundary conditions on the boundary of the disk. The proof goes through unchanged in the case of the plane by taking $\ell\to\infty$, with $(\ell\uD)_{\ell\to\infty}$ identified with the plane $\C$.

Consider a product of local fields $\prod_j \Or_j(x_j)$ for disjoint coordinates $x_j\in\ell\uD\setminus \{0\}$ lying in the disk minus the origin $(0,\ell)$ (see the paragraph at the end of subsection \ref{ssectsingle}). We now show \eqref{tman}.

The manifold structure on which the QFT is placed may be extracted by analyzing the smooth paths obtained by (Euclidean) space-time translations of local fields. Smooth paths are naturally produced by independent smooth variations of the $x_j$ coordinates above. The structure of the manifold is that arising from smoothness of correlation functions along smooth paths. Correlation functions in a QFT on a manifold ${\cal M}$ are Euclidean tensor fields on the manifold ${\cal M}'$ which is the direct product of many copies of ${\cal M}$ (as many copies as there are fields involved in the correlation function) from which all diagonals have been removed (avoiding colliding positions of fields). The tensorial properties of correlation functions are determined by Euclidean transformation properties of the local fields involved. Euclidean transformation properties are sufficient to unambiguously define correlation functions on any manifold with Euclidean transition function (flat manifold).

Consider the left-hand side of \eqref{tman}. Consider for now $x_j\not\in(0,\ell)$. By locality and the fact that, according to \eqref{renang}, the conical field is supported on $[0,\ell)$, the correlation function is smooth along an $x_j$-path whenever the half-segment $[0,\ell)$ is not intersected by the path\footnote{Correlation functions are defined and finite for any local field insertions at separate positions. In space-time translation invariant QFT, any derivative of a field, of any order,  is associated with a local field, and thus smoothness is guaranteed. In the quantization scheme on the line, the space of all local fields is generated, from a possibly infinite generating set, by taking space derivatives, and time derivatives are obtained by commuting with the Hamiltonian and thus expressed in terms of local fields. The resulting expressions are local and independent of the quantization scheme.}. Hence the correlation function is smooth on $\uD\setminus[0,\ell)$, which we identify with $\check\uD^{(1)}$.

In the expression \eqref{renang2}, one can define, in a similar way as that done in \eqref{regang} and \eqref{renang2}, a renormalized exponential involving any smooth integration path $\gamma$ from $0$ to the boundary of the disk $\ell S^1 = \{z:|z|=\ell\}$, in place of the path lying on $[0,\ell)$. The angle-splitting procedure is replaced by any splitting within a tubular neighborhood of $\gamma$. Since the rotation current $\frak R^\mu(x)$ is a conserved Noether current and thanks to the rotation-preserving boundary conditions, by the Ward-Takahashi identities the zero-splitting limit exists. Further, any two homotopic paths $\gamma$ on the topological space $\check\uD\setminus \cup_j\{x_j\}$ will lead to the same correlation function \eqref{tman}, and any two parts of $\gamma$ that trace the same segment of curve but in opposite directions can be deleted. In the wedge bounded by $e^{-\ri \alpha}[0,\ell)$ and $[0,\ell)$ in $\ell\uD$, we use the transition function $U^{10}$ in order to interpret the fields as lying in $\check\uD^{(0)}$ (see the end of subsection \ref{ssectsingle}). But using such deformations and deletions, it is then possible to extend smoothly the correlation function from that wedge across the half-segment $[0,\ell)$: local fields on $(0,\ell)$ lie in $\check\uD^{(0)}_{(-\alpha,\alpha)}$, and those lying in the wedge bounded by $[0,\ell)$ and $e^{\ri\alpha}[0,\ell)$ are dressed by being encircled by small closed paths $\gamma$. We interpret these dressed fields as lying in $\check\uD^{(0)}_{(0,\alpha)}$. This gives a smooth extension.

Since $\frak R^\mu(x)$ is the Noether current associated with rotations, the modification produced by encircling a local field by a small closed path $\gamma$ is that of a rotation by an angle $-\alpha$ (rotation clockwise by $\alpha$) with respect to the origin. We may interpret the resulting fields as lying in $\check\uD^{(2)}$, and the rotation implements the transition function $U^{(02)}$ (a rotation) according to the appropriate tensor field property. We may then move the encircled fields along smooth paths going to angles greater than $\alpha$, beyond the wedge bounded by $[0,\ell)$ and $e^{\ri\alpha}[0,\ell)$ in $\uD$, thus smoothly extending the correlation function to that region. Rotation now brings the fields into $\check\uD^{(1)}$: the corresponding part of $\check\uD^{(2)}$ is overlapping with $\check\uD^{(1)}$, with transition function $U^{(21)}$. This completes the proof that the left-hand side of \eqref{tman} is the tensor field on the manifold ${\cal M}_{\alpha,\ell}$ given by the right-hand side.

The proof above is general. In the case of CFT, we could give an alternative proof by using the fact that ${\cal M}_{\alpha,\ell}$ is conformally equivalent to the punctured disk, via the transformation ${\cal M}_{\alpha,\ell} \to \ell^{1/n}\uD\setminus \{0\}$, $z\mapsto y = z^{1/(1+\alpha/2\pi)} = z^{1/n}$. This follows the lines of the previous subsection, using angular quantization and combining the maps involved up to the final space, the punctured disk parametrized by the variable $y$ (see just above \eqref{ft}). The field $\cV_\alpha$ provides an explicit implementation of this conformal mapping $z\mapsto z^{1/n}$ on correlation functions, via an exponential of the stress-energy tensor.

\subsection{Logarithmic fields} \label{sssectlog}

Property III is more subtle, and here we show it by studying a certain CFT field with logarithmic properties.

The main idea is that, since the limit in $\varep$ exists (property I), then we only need to show that the regularized exponential transforms trivially. The scaling property of the renormalized exponential then follows from the power of $\varep$.

Recall that the rotation current in \eqref{r0} and \eqref{r0cft} involves not only the stress-energy tensor components, but also a term proportional to the identity operator ${\bf 1}$. This term of course does not affect the manifold construction. It corresponds to an overall normalization factor of the renormalized exponential, and in particular the observed singularity as $z\to0$ amounts to a modification of the power of $\varep$ required in the renormalized exponential \eqref{renang} and \eqref{renang2} in order to make the limit exist. Any different choice of proportionality factor in place of $c/12$ would lead to a similar construction with a different power of $\varep$, and likewise the result \eqref{norm} with a different power of $\ell$. This singularity is fixed solely by the condition that the regularized exponential transforms trivially.

In order to establish trivial transformation of the regularized exponential, we need a more accurate analysis. Recall that local QFT fields can be classified according to their properties under local scale transformation, $\Or(x)\mapsto \Or^{(\lambda)}(x)$ where $\lambda$ is the scale factor. In general QFT, this determines how the renormalization group acts on the field.  The conical field $\cV_\alpha(0)$ is the renormalized exponential $e^{(\alpha/2\pi) h(0)}$ of the nonlocal field
\[
	h(0) = \int_0^\infty \dd \rx \, j(\rx),\quad j(\rx) = -2\pi\frak{R}^2(\rx)
\]
(here for simplicity on the plane and as an integration along the $\rx$ direction). Write the dilation generator as ${\cal D} = \int \dd\rx\, \frak{D}(x)$. Clearly, on the part of the branch that lies away from 0, it acts as $\int \dd\rx\,[{\cal D},j(\rx)]$. However, around 0 the scale transformation is more subtle. It is sufficient to analyze its effect on the local neighbourhood around 0. For this purpose, we consider a CFT, and we take the dilation operator to lie within a disk of radius $\ell$. Since dilations do not preserve the disk, terms will appear, in the infinitesimal scale transformation $\ad {\cal D}(h(0))$, that lie at $\ell$. These however should be interpreted as contributing to the large-distance transformation properties, beyond the CFT description, and must be discarded in the assessment of how dilation acts on the local neighbourhood. We therefore propose that, within the disk of radius $\ell$, the infinitesimal scale transformation can be determined by
\beq\label{presc}
	\PP_{(-\ell,\ell)} \,\ad {\cal D}(h(0)) =  
	\PP_{(-\ell,\ell)} \int_{-\ell}^\ell \dd\rx\,[\frak{D}(\rx),h(0)] \qquad \mbox{(CFT)}
\eeq
where $h(0)$ extends beyond $\ell$, and where $\PP_{(-\ell,\ell)}$ projects onto the interval $(-\ell,\ell)$ (discarding local fields at radius $\ell$ or beyond).

In effect, this prescription accounts for the effects of the dilation on the UV boundary around $0$ in the regularized exponential, but not for those of any IR boundary, thus describing the local scaling properties of the renormalized exponential.

We will show that under this scale transformation, the regularized exponential in \eqref{renang} (or \eqref{renang2}) transforms {\em trivially}: it is only affected by a scale transformation of $\varep$,
\beq\label{trivsc}
	\lt[e^{- \alpha\int_0 \dd x^\mu \ep_{\mu\nu}\frak R^\nu(x)}\rt]_\varep^{(\lambda)}
	=
	\lt[e^{- \alpha\int_0 \dd x^\mu \ep_{\mu\nu}\frak R^\nu(x)}\rt]_{\lambda\varep}.
\eeq
The existence of the limit proved above then guarantees that the renormalized exponential has scaling dimension $2\Delta_n$.

Consider the holomorphic field, here for convenience at more general positions $z$,
\beq\label{h0}
	h(z) = \int_{z}^{\ell} \dd s\,(s-z)\,T(s)
\eeq
and its conjugate $\b h(\b z)$ (anti-holomorphic). We are interested in the regularized exponential
\beq\label{cthe}
	\lt[e^{\frc{\alpha}{2\pi} \lt(h(0) + \b h(0) - \frc{c{\bf 1}}{12}
	\int_0^\ell \lt(\frc{\dd z}z+ \frc{\dd \b z}{\b z}\rt)\rt)}\rt]_\varep
	=\lt[e^{\gamma (h(0)+ \b h(0))}\rt]_\varep
	\lt[e^{-\frc{c \gamma {\bf 1}}{12}
	\int_0^\ell \lt(\frc{\dd z}z+ \frc{\dd \b z}{\b z}\rt)}\rt]_\varep
\eeq
where $\gamma = \frc{\alpha}{2\pi}$. The trivial scaling of the regularized exponential follows from the {\em logarithmic transformation properties} of the fields $h_0(z)$ and $\b h_0(\b z)$:
\beq\label{logprop}
	h^{(\lambda)}(z) = h(\lambda z) + \frc{c}{12} \log \lambda\,{\bf 1},\quad
	\b h^{(\lambda)}(\b z) = \b h(\lambda \b z) + \frc{c}{12} \log \lambda\,{\bf 1}.
\eeq
That is, $h(z)$ is a zero-dimensional logarithmic field with respect to scaling transformations, with logarithmic partner the identity field ${\bf 1}$,
\beq\label{trianl0}
	\ad{\cal D} \mato{c} {\bf 1} \\ h\matf
	= \mato{cc} 0&0 \\
	c/12 & 0 \matf  \mato{c} {\bf 1} \\ h\matf,
\eeq
and similarly for $\b h(\b z)$. Taking into account that the scale transformation operator $\ad{\cal D}$ acts on the hole by scaling its radius, equation \eqref{logprop} implies
\beq\label{alphah}
	\lt[e^{\gamma (h(0)+\b h(0))}\rt]_\varep^{(\lambda)} =
	\lambda^{\gamma \frc{c}{6}}
	\lt[e^{\gamma (h(0)+\b h(0))}\rt]_{\lambda \varep}.
\eeq
With the immediate relation
\beq\label{imme}
	\lt[e^{-\frc{c\gamma {\bf 1}}{12}
	\int_0^\ell (\dd z/z+ \dd \b z/ \b z)}\rt]_{\lambda\varep}
	= \lambda^{\gamma \frc{c}{6}}
	\lt[e^{-\frc{c\gamma {\bf 1}}{12}
	\int_0^\ell (\dd z/z+ \dd \b z/ \b z)}\rt]_{\varep},
\eeq
equation \eqref{trivsc} indeed follows.

We show \eqref{logprop} as follows.  We use scaling operator ${\cal D} = \cL_0 + \b\cL_0$, with
\beq\label{L0}
	\cL_{0} = \frc{\ii}{2\pi} \int \dd x\,x T(x),\quad
	\bar{\cL}_{0} = -\frc{\ii}{2\pi} \int \dd x\,x \b T(x).
\eeq
According to the prescription \eqref{presc} (translated to the point $z$), we must evaluate
\[
	\frc{\ii}{2\pi}
	\; \PP_{(z-\ell,z+\ell)}\int_{z-\ell}^{z+\ell} \dd x\,[xT(x),h(z)].
\]
The commutator involved can be calculated using
\beq\label{commut}
	\frc{\ii}{2\pi} [T(x),T(z)] = -\frc{c}{12} \delta'''(x-z)\,{\bf 1}
	- 2 \delta'(x-z)\,T(z)
	+ \delta(x-z)\,\p T(z)
\eeq
which follows from the standard OPE
\beq\label{TTz}
	T(x)T(z) 
	\sim \frc{c\,{\bf 1}}{2(x-z)^4}
	+ \frc{2\,T(z)}{(x-z)^2} + \frc{\p T(z)}{x-z} + {\rm reg.}
\eeq
We find
\beqa
	\lefteqn{\frc{\ii}{2\pi} [T(x),h(z)] = \frc{\ii}{2\pi}
	\int_z^\infty\dd s\,(s-z) [T(x),T(s)]} &&\n
	&=&
	\lt(-\frc{c}{12} \lt(\frc{\p}{\p x}\rt)^3 \big((x-z)\,\Theta(x-z)\big){\bf 1}
	- 2\frc{\p}{\p x} \big((x-z)\Theta(x-z)\,T(x)\big)
	+ (x-z)\Theta(x-z)\,\p T(x)\rt)\no
\eeqa
where $\Theta(x)$ is Heaviside's step function. Applying the integral $\int_{-\ell}^{\ell} dx\,x$ and using integration by parts, we obtain
\beq
	\frc{\ii}{2\pi}\int_{z-\ell}^{z+\ell} \dd x\,x [T(x),h(z)]
	= \frc{c}{12}  {\bf 1}
	- z\,\int_{z}^{z+\ell} \dd x\,T(x) - (z+\ell)\ell\,T(z+\ell).
\eeq
Applying $\PP_{(z-\ell,z+\ell)}$ the last term drops. This finally gives
\beq\label{adl0}
	\PP_{(z-\ell,z+\ell)}\,\ad{\cal D}(h(z))
	= \PP_{(z-\ell,z+\ell)}\lt(\frc{c}{12}  {\bf 1}
	+ z\,\p h(z)\rt).
\eeq
Extending beyond the local neighbourhood of $z$ described by CFT, we therefore conclude that $\ad{\cal D}(h(z)) = \frc{c}{12}  {\bf 1}+ z\,\p h(z)$. A similar result holds for $\b h(\b z)$. Exponentiating, this gives \eqref{logprop}.


\end{document}